\documentclass[aps,pra,reprint,superscriptaddress,letterpaper,twocolumn,floatfix]{revtex4-2}

\usepackage{amsmath,amssymb,mathtools,dsfont,esint,textgreek,soul}

\usepackage{graphicx,setspace}
\usepackage[usenames, dvipsnames, x11names]{xcolor}
\usepackage[colorlinks]{hyperref}
\usepackage{CJKutf8}

\hypersetup{
 colorlinks=true,
 citecolor=red,
 linkcolor=blue,
 urlcolor=cyan}

\newcommand{\proj}{\hat{\mathcal{P}}_{s}}
\newcommand{\drive}{\tilde{\Omega}}
\newcommand{\sqr}{\mathrel{\raisebox{0.09ex}{${\scriptscriptstyle\blacksquare}$}}}
\newcommand{\Jbar}{\bar{J}}
\begin{document}
\begin{CJK}{UTF8}{gbsn}

\title{Exact Results for a Boundary-Driven Double Spin Chain and Resource-Efficient Remote Entanglement Stabilization}

\date{\today}
\author{Andrew Lingenfelter}
\email{lingenfelter@uchicago.edu}
\affiliation{Pritzker School of Molecular Engineering, University of Chicago, Chicago, IL 60637, USA}
\affiliation{Department of Physics, University of Chicago, Chicago, IL 60637, USA}
\author{Mingxing Yao}
\affiliation{Pritzker School of Molecular Engineering, University of Chicago, Chicago, IL 60637, USA}
\author{Andrew Pocklington}
\affiliation{Pritzker School of Molecular Engineering, University of Chicago, Chicago, IL 60637, USA}
\affiliation{Department of Physics, University of Chicago, Chicago, IL 60637, USA}
\author{Yu-Xin Wang (王语馨)}
\affiliation{Pritzker School of Molecular Engineering, University of Chicago, Chicago, IL 60637, USA}
\author{Abdullah Irfan}
\affiliation{Department of Physics, University of Illinois at Urbana-Champaign, Urbana, IL 61801, USA}
\author{Wolfgang Pfaff}
\affiliation{Department of Physics, University of Illinois at Urbana-Champaign, Urbana, IL 61801, USA}
\author{Aashish A. Clerk}
\affiliation{Pritzker School of Molecular Engineering, University of Chicago, Chicago, IL 60637, USA}

\begin{abstract}

We derive an exact solution for the steady state of a setup where two XX-coupled $N$-qubit spin chains (with possibly non-uniform couplings) are subject to boundary Rabi drives, and common boundary loss generated by a waveguide (either bidirectional or unidirectional).  For a wide range of parameters, this system has a pure entangled steady state, providing a means for stabilizing remote multi-qubit entanglement without 
the use of squeezed light.  Our solution also provides insights into a single boundary-driven dissipative XX spin chain that maps to an interacting fermionic model.  The non-equilibrium steady state exhibits surprising correlation effects, including an emergent pairing of hole excitations that arises from dynamically constrained hopping.  Our system could be implemented in a number of experimental platforms, including circuit QED.  
\end{abstract}

\maketitle
\end{CJK}

\section{Introduction}
\label{sec:intro}

The recent intense interest in driven-dissipative quantum systems has many distinct motivations.  One key goal is to understand how tailored dissipative processes \cite{poyatos_Quantum_1996,plenio_Entangled_2002} could be used to stabilize entangled quantum states in both the few and many-body regime, with possible applications to quantum information processing (see e.g.~\cite{kraus_Discrete_2004,stannigel_Drivendissipative_2012,schirmer_Stabilizing_2010,motzoi_Backactiondriven_2016,govia_Stabilizing_2022,brown_Trade_2022,doucet_High_2020,vollbrecht_Entanglement_2011,ma_Stabilizing_2019,ma_Couplingmodulation_2021,agusti_Longdistance_2022,horn_Quantum_2018,diehl_Quantum_2008,pocklington_Stabilizing_2022,zippilli_Steadystate_2015,angeletti_Dissipative_2023,zippilli_Entanglement_2013,yanay_Reservoir_2018,zippilli_Dissipative_2021,ma_Pure_2017}).  A second, seemingly distinct line of work seeks to understand the unique properties of non-equilibrium steady states (NESS) that arise in driven quantum spin chains from the interplay of driving, lattice dynamics and dissipation \cite{znidaric_Exact_2010,prosen_Third_2008,sigurdsson_Drivendissipative_2017,popkov_Exact_2019,yamanaka_Exact_2023,znidaric_Manybody_2008,znidaric_Relaxation_2015,medvedyeva_Powerlaw_2014,prosen_Quantum_2008,prosen_Matrix_2009,prosen_Open_2011,parmee_Steady_2020,vitagliano_Volumelaw_2010}.  Here, certain exact solutions have been especially valuable \cite{znidaric_Exact_2010,prosen_Third_2008,sigurdsson_Drivendissipative_2017,popkov_Exact_2019,yamanaka_Exact_2023}.

\begin{figure}[t!]
     \centering
    \includegraphics[width=0.98\columnwidth]{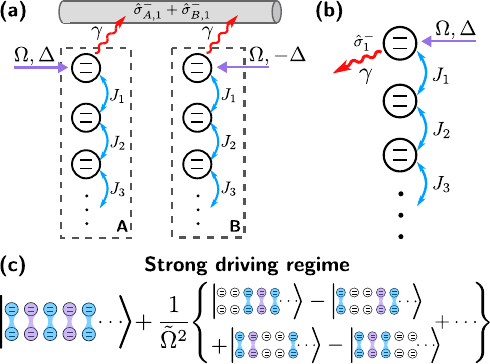}
     \caption{
        \textbf{(a)} Two XX-coupled $N$ qubit chains, with boundary collective loss (rate $\gamma$, mediated by a waveguide) and Rabi drives (strength $\Omega$, detuning $\pm\Delta$).
        \textbf{(b)} A single XX chain with a Rabi drive and loss on the boundary, which corresponds to an interacting fermionic model. 
        The steady state here is obtained from the pure steady state of the two-chain system.
        \textbf{(c)} For strong dimensionless driving $\drive$ (cf. Eq.~\eqref{eq:eff-drive}), the steady state approaches a product of dimer Bell pairs.  For finite driving, the exact steady state is obtained by doping this state with delocalized, paired ``holes".  Strong inter-chain entanglement can be achieved even if $\Omega < \gamma$.  
    }
     \label{fig:intro}
 \end{figure}

In this work, we present new exact results for two boundary-driven spin models that are directly relevant to both of the above motivations.  The first (Fig.~\ref{fig:intro}(a)) consists of two passively-coupled $N$-qubit chains that hang off the same waveguide.  We show that for arbitrary $N$, this system has a pure, highly-entangled steady, even for weak driving and with certain kinds of disorder.  The second 
(Fig.~\ref{fig:intro}(b)) is a single qubit chain with boundary Rabi driving and loss, which somewhat surprisingly corresponds to an {\it interacting} fermionic model.  We nonetheless obtain an exact result for the NESS by considering the directional waveguide limit of the double-chain system: the double-chain pure state is the purification of the desired NESS. This represents the first use of the hidden time-reversal / quantum absorber exact solution method \cite{stannigel_Drivendissipative_2012,roberts_DrivenDissipative_2020,roberts_Hidden_2021} to a non-trivial system where interactions are not long-ranged.     
This solution is also a rare example of an exactly solved coherently-driven 1D spin chain model.  In contrast, the few existing examples of exactly solvable boundary-driven spin chains have involved purely incoherent drives \cite{znidaric_Exact_2010,prosen_Third_2008,sigurdsson_Drivendissipative_2017,popkov_Exact_2019,yamanaka_Exact_2023}.

%%%%%%%%%%%%%%%%%%%%%%%%%%%%%%%%%%%%%%%
%%%%%%%%%%%%%%%%%%%%%%%%%%%%%%%%%%%%%%%

Our exact solutions provide a wealth of insights relevant to understanding correlations in the NESS, and to the application of remote entanglement stabilization. Despite the lack of any explicit attractive interactions in our systems, their steady states exhibit strong real-space pairing correlations.  In fact, we show that the pure steady state of the double-chain system can be exactly written as a condensate of paired holes, where a hole here corresponds to an interchain dimer of qubits that are both in the vacuum state
(see Fig.~\ref{fig:intro}(c)).  We discuss how this pairing has clear observable consequences, and how it ultimately arises from a kinetically constrained hopping process (something that could also be studied in certain non-dissipative cold atom systems \cite{mamaev_Quantum_2019}). 
This pairing mechanism also has a direct connection to the emergence of quantum scar states: it represents a restricted spectrum generating algebra of the two-chain Hamiltonian \cite{moudgalya_etapairing_2020}, and can be used to construct a tower of many-body scar states in related nonintegrable ladder models such as that studied in Ref.~\cite{znidaric_Coexistence_2013}.
We also discuss regimes where this pairing structure directly results in an NESS with features reminiscent of charge density wave order.

In terms of entanglement stabilization, our double-chain system has potential advantages over other proposals \cite{yanay_Reservoir_2018,pocklington_Stabilizing_2022,zippilli_Dissipative_2021,zippilli_Entanglement_2013,zippilli_Steadystate_2015,angeletti_Dissipative_2023} as it does not require the preparation and transport of high-fidelity squeezed light, but only simple Rabi drives and passive waveguide couplings.  It unifies and extends to the multi-qubit regime previously known two-qubit schemes \cite{schirmer_Stabilizing_2010,stannigel_Drivendissipative_2012,motzoi_Backactiondriven_2016,govia_Stabilizing_2022,brown_Trade_2022}.  For large Rabi drive amplitude $\Omega$, our system is just one example of a more general mechanism for replicating definite-parity two qubit states using simple XX (hopping) couplings.  While this was seen previously in different systems \cite{pocklington_Stabilizing_2022,zippilli_Entanglement_2013,zippilli_Steadystate_2015,angeletti_Dissipative_2023}, the underlying mechanism had not been elucidated.  For finite $\Omega$, the two qubit systems need $\Omega > \gamma$ for significant steady state entanglement, where $\gamma$ is the waveguide-induced dissipation.  We find that this is no longer the case for larger systems:  for $N \geq 2$, strong steady state entanglement only requires the potentially much weaker condition $\Omega > \sqrt{J \gamma}$, where $J$ is the hopping (XX coupling) in each chain.  We also show that the method introduced in Ref.~\cite{brown_Trade_2022} for speeding up the dissipative stabilization of a two-qubit entangled state can be extended to situations with multiple qubits, leading to a dramatic acceleration of our protocol.  

The rest of this paper is organized as follows.  In Sec.~\ref{sec:setup} we introduce our two basic models, while in Sec.~\ref{sec:steady-state} we give the exact pure steady state of the double chain,  discuss its structure, and explain the general \emph{entanglement replication} mechanism that applies for large $\Omega$. 
In Sec.~\ref{sec:hole-pairing} we explain three surprising features of the steady state: effective hole-pairing, universal single-parameter scaling of the excitation density, and the emergence of single particle states with charge density wave order.
In Sec.~\ref{sec:entanglement} we discuss the application of Fig.~\ref{fig:intro}(a) to remote many-Bell-pair entanglement stabilization.

%%%%%%%%%%%%%%%%%%%%%%%%%%%%%%%%%%%%%%%%%%%%%%%%%%%%%%%%%%%%%%%%%%%%%%
%%%%%%%%%%%%%%%%%%%%%%%%%%%%%%%%%%%%%%%%%%%%%%%%%%%%%%%%%%%%%%%%%%%%%%

\section{{\it XX}-coupled qubit chains with boundary dissipation and driving}
\label{sec:setup}

\subsection{The two-chain model}

Consider the setup in Fig.~\ref{fig:intro}(a): two passively coupled $N$-qubit chains ($A$ and $B$), with boundary driving and correlated dissipation.  
The dynamics is described by the Lindblad master equation
\begin{align}
    \partial_t \hat{\rho} &= -i[\hat{H},\hat{\rho}] + \gamma\mathcal{D}[\hat{c}]\hat{\rho} \label{eq:qme}\\ 
    \hat{H} &= \hat{H}_{\rm drive} + \hat{H}_{\rm XX} + \hat{H}_{\rm diss}, \label{eq:H} \\    
    \hat{c} &= \hat{\sigma}_{A,1}^{-} + \hat{\sigma}_{B,1}^{-},\label{eq:closs}
\end{align}
where the Hamiltonian terms are given by
\begin{align}
    \hat{H}_{\rm drive} &= \frac{\Omega}{2}\left( \hat{\sigma}_{A,1}^{x} + \hat{\sigma}_{B,1}^{x}\right) + \frac{\Delta}{2}\left( \hat{\sigma}_{A,1}^z - \hat{\sigma}_{B,1}^z \right),
    \label{eq:Hdrive}\\
    \hat{H}_{\rm XX} &= \frac{1}{2} \sum_{j=1}^{N-1}\sum_{s=A,B}J_j\left(\hat{\sigma}_{s,j}^{+}\hat{\sigma}_{s,j+1}^{-} + {\rm h.c.} \right).\label{eq:Hxx} \\
    \hat{H}_{\rm diss} &= \frac{1}{2}i\nu\gamma\left(\hat{\sigma}_{A,1}^{+}\hat{\sigma}_{B,1}^{-} - {\rm h.c.}\right).
    \label{eq:Hdiss}
\end{align}
The Lindblad dissipator $\mathcal{D}[\hat{c}]\,\cdot = \hat{c}\,\cdot\,\hat{c}^\dagger - \{\hat{c}^\dagger\hat{c},\,\cdot\,\}/2$ describes collective loss on the site 1 qubits (with  $\hat{\sigma}^- = |0\rangle\langle 1|$,  $\hat{\sigma}^z = |1\rangle\langle 1| - |0\rangle\langle 0|$).

Consider first the driving of our system.  Qubits $A1$ and $B1$ are Rabi-driven at the same frequency and amplitude $\Omega$.  We however take qubit $A1$ ($B1$) to be detuned by $+\Delta$ ($-\Delta$) from the drive frequency.  Treating these drives within the rotating wave approximation, we obtain the rotating frame Hamiltonian $\hat{H}_{\rm drive}$, given by Eq.~\eqref{eq:Hdrive}.
We will take the remaining qubits in each chain to be resonant with the drive frequency.  

Within each chain, excitations can hop between adjacent qubits.  This is described by simple nearest-neighbor XX couplings: $\hat{H}_{\rm XX}$, given by Eq.~\eqref{eq:Hxx}.
While the hopping amplitudes $J_j$ in each chain can vary from bond to bond, we require that the hopping across a particular bond $j$ is the same for chain $A$ and $B$; as we will see in Sec.~\ref{sec:steady-state}, this mirror symmetry is necessary to obtain a pure steady state. 
The passive exchange couplings we use here are natural in many experimental settings.  For example, in superconducting circuits
they could be realized straightforwardly with capacitive couplings.

Finally, we turn to the dissipation-mediated coupling between the two chains.  Qubits $A1$ and $B1$ experience collective loss (at rate $\gamma$) due to a common coupling to a Markovian reservoir.  We focus on the case where this bath is an open waveguide structure, and the two chains are spatially separated (in the limit where non-Markovian effects associated with a finite propagation time can be neglected).  We consider two types of waveguides: (i) a bidirectional waveguide that supports both left- and right-propagating waves, or (ii) a unidirectional waveguide that supports only, e.g., right-propagating waves.
Note that a bidirectional waveguide requires precise spacing of the qubits to engineer collective loss (see e.g.~Ref.~\cite{govia_Stabilizing_2022}); such control is not needed for the directional setup.
When the waveguide is not fully bidirectional (e.g.~qubits couple preferentially to right-propagating modes versus left-propagating modes), it induces the effective exchange Hamiltonian $\hat{H}_{\rm diss}$ given by Eq.~\eqref{eq:Hdiss}~\cite{metelmann_Nonreciprocal_2015},
where $\nu$ is a directionality factor, $-1\leq\nu\leq 1$.
When $\nu=0$ the waveguide is perfectly bidirectional and when $\nu=+1$ ($\nu=-1$) the waveguide is perfectly unidirectional with system $A$ (system $B$) upstream of $B$ ($A$).

\subsection{Remote two-qubit entanglement stabilization}
\label{subsec:setup-many-qb-entanglemnt-stabilization}

A key motivation for our two chain setup is the ability to stabilize large amounts of {\it remote} entanglement.  To understand the challenge here, we first review the simpler problem of dissipatively stabilizing entanglement between two remote qubits.
One generic approach is to use squeezed light (as first introduced by Kraus et al \cite{kraus_Discrete_2004}, and further studied in 
\cite{vollbrecht_Entanglement_2011,agusti_Longdistance_2022,angeletti_Dissipative_2023}).  Such schemes ultimately rely on generating correlated ``pairing" dissipation, with a Lindblad jump operator
$\hat{c}_{\rm pair} = u\hat{\sigma}_{A,1}^- + v\hat{\sigma}_{B,1}^+$.
 While conceptually appealing, such pairing-based protocols are experimentally challenging, given the difficulty of preparing and propagating high quality squeezed light.
Recent work shows that pairing dissipation can be realized without squeezed light, instead using modulated qubit-waveguide couplings~\cite{govia_Stabilizing_2022,pocklington_Stabilizing_2022}; this is also challenging in many setups.  
 
A simpler approach for stabilizing remote two qubit entanglement (using only Rabi drives and passive loss) is provided by the $N=1$ version of Eq.~\eqref{eq:qme}.
This both unifies and generalizes the entanglement stabilization schemes studied in Refs.~\cite{schirmer_Stabilizing_2010,stannigel_Drivendissipative_2012,motzoi_Backactiondriven_2016,govia_Stabilizing_2022}.
The bidirectional waveguide case ($\nu=0$) yields the schemes of Refs.~\cite{motzoi_Backactiondriven_2016,govia_Stabilizing_2022} and the unidirectional waveguide case ($\nu=\pm1$) yields the scheme introduced in Ref.~\cite{stannigel_Drivendissipative_2012}.

The general $N=1$ system has a pure steady state $\hat{\rho}_1(t\to\infty) = |\psi_1\rangle\langle\psi_1|$ given by (up to normalization)
\begin{align}
    |\psi_1\rangle &= \sqrt{2}\Gamma|00\rangle + \Omega|S\rangle, \label{eq:steady-state-2qb}\\
    \Gamma &\equiv \Delta - \frac{1}{2}i\nu\gamma, \label{eq:Gamma}
\end{align}
where $\Gamma\in\mathbb{C}$ is a generalized complex detuning.
Here we introduce the singlet and triplet entangled states
\begin{align}
    |S\rangle &= \frac{1}{\sqrt{2}}(|01\rangle-|10\rangle),\quad
    |T\rangle = \frac{1}{\sqrt{2}}(|01\rangle+|10\rangle), \label{eq:dimer-bell-states}
\end{align}
in addition to the unentangled vacuum $|00\rangle$ and the ``doublon'' $|11\rangle$.
Eq.~\eqref{eq:steady-state-2qb} is the \emph{unique} two-qubit steady state for any $|\Omega/\Gamma|<\infty$, and as $|\Omega/\Gamma|\to\infty$, it approaches a perfect Bell state $|\psi_1\rangle \to |S\rangle$.
Note, however, that in this limit the dissipative gap closes and a second \emph{impure} steady state emerges \cite{doucet_High_2020,brown_Trade_2022,govia_Stabilizing_2022}.

Going forward, one key goal will be to extend this entanglement stabilization to the case where each chain has $N \gg 1$ qubits.  As we show, this is a priori a non-trivial exercise. Unlike schemes that use squeezing dissipation, the entanglement structure of the $N \gg 1$ chain is complicated and crucially depends on the interplay between the boundary driving and lattice dynamics. However, this interplay also gives the $N\gg 1$ chain advantages over squeezing based schemes by enabling large amounts of entanglement to be stabilized without requiring $\Omega \gg \gamma$. To describe the stabilized state for $N\gg 1$, it will be useful to use a basis where we specify the state of each cross chain dimer $j$.  We will use the notation like $|(01)_j\rangle=|0_{A,j}1_{B,j}\rangle$ to denote the state of the qubit pair on dimer site $j$.  For example, $|S_j\rangle$ denotes a Bell pair spanning qubits $Aj$ and $Bj$.

\subsection{Boundary driven dissipative quantum spin chain}
\label{subsec:boundary-diss-spin-chain}

A second motivation for our work comes from the seemingly simpler single-chain system depicted in Fig.~\ref{fig:intro}(b):   a chain of XX-coupled qubits with local loss and Rabi driving on one boundary site of the chain.  The master equation for this system is
\begin{align}
    \partial_t \hat{\rho}_A &= -i[\hat{H}_A,\hat{\rho}_A] + \gamma\mathcal{D}[\hat{\sigma}_{A,1}^-]\hat{\rho}_A, \label{eq:single-chain-qme}\\
    \hat{H}_A &= \frac{\Omega}{2}\hat{\sigma}_{A,1}^x + \frac{\Delta}{2}\hat{\sigma}_{A,1}^z + \frac{1}{2}\sum_j J_j\big( \hat{\sigma}_{A,j}^+ \hat{\sigma}_{A,j+1}^- + {\rm h.c.} \big).\nonumber
\end{align}
As with other boundary-driven spin chain systems, we are interested in understanding the NESS of this setup.  For weak drives one expects the NESS approaches a product state (all qubits in 
$|0\rangle$), whereas for strong driving, one instead expects an infinite temperature state.  How one interpolates between these limits (and how the corresponding NESS depends on master equation parameters and the disordered hopping) is at first glance unclear.  Surprisingly, Eq.~\eqref{eq:single-chain-qme} {\it cannot} be mapped to a system of free fermions, even though this is possible for the Hamiltonian $\hat{H}_A$ alone.
As we show in App.~\ref{app:single-spin-chain}, upon making a Jordan-Wigner transformation, the Rabi drive in $\hat{H}_A$ yields a linear-in-fermions term, something that can be treated using the method of Ref.~\cite{colpa_Diagonalisation_1979}.  However, when applied to the full master equation, this necessarily yields an interacting fermionic problem. Specifically, the loss dissipator becomes nonlinear in the fermionic master equation: $\partial_t \hat\rho = -i[\hat H_{{\rm fermi},\eta},\hat\rho] + \gamma\mathcal{D}[(-1)^{\hat\eta^\dagger\hat\eta}\hat c_1]\hat\rho$, where $\hat H_{{\rm fermi},\eta}$ is a quadratic fermionic Hamiltonian, $\hat c_1$ is the fermion lowering operator on the dissipative site, and $\hat\eta$ is the fermionic lowering operator of an auxiliary mode introduced per the method of Ref.~\cite{colpa_Diagonalisation_1979} (see App.~\ref{app:single-spin-chain} for details).

Eq.~\eqref{eq:single-chain-qme} thus corresponds to an interacting fermionic system without translational invariance (the hoppings can be disordered).  Surprisingly, we are able to find an exact analytic description of its NESS.  We do this by first solving for the pure steady state $|\psi\rangle$ of the double-chain system in Eq.~\eqref{eq:qme}, something that can be be done analytically.  If we then focus on the case where the waveguide is directional from $A$ to $B$ (i.e. $\nu = 1$), simply tracing out the $B$ chain yields the steady state of the single-chain system in Eq.~\eqref{eq:single-chain-qme}.
\begin{align}
    \hat{\rho}_A = {\rm tr}_B |\psi\rangle\langle\psi|,
\end{align}
This corresponds to a new many-body application of the coherent quantum absorber technique introduced in Ref.~\cite{stannigel_Drivendissipative_2012} and extended in Refs.~\cite{roberts_Hidden_2021,roberts_DrivenDissipative_2020,roberts_Competition_2023,roberts_Exact_2023}.  The existence of this exact solution implies that the single-chain system has a ``hidden time reversal symmetry" which enforces Onsager time-symmetry of a certain class of two-time correlation functions \cite{roberts_Hidden_2021}.

As presented in more detail in Sec.~\ref{sec:hole-pairing}, our exact solution for this boundary-driven spin chain reveals a number of surprising features in the NESS, including regimes of strong long-range correlations and even structures reminiscent of charge density wave order.

%%%%%%%%%%%%%%%%%%%%%%%%%%%%%%%%%%%%%%%%%%%%%%%%%%%%%%%%%%%%%%%%%%%%%%
%%%%%%%%%%%%%%%%%%%%%%%%%%%%%%%%%%%%%%%%%%%%%%%%%%%%%%%%%%%%%%%%%%%%%%

\section{Pure entangled steady state for arbitrary \texorpdfstring{$N$}{N} and \texorpdfstring{$\Omega$}{\textOmega}}
\label{sec:steady-state}

\begin{figure}[t!]
     \centering
    \includegraphics[width=0.98\columnwidth]{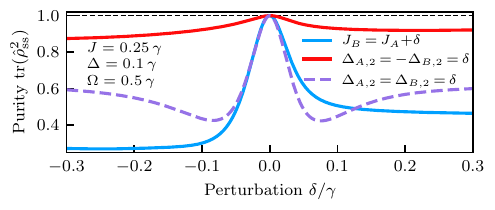}
     \caption{
        \textbf{The existence of a pure steady state is special.}
        We consider two kinds of Hamiltonian perturbations to the $N=2$ version of the system in Fig.~\ref{fig:intro}(a), demonstrating that the emergence of a pure steady state is not generic.  
        We break the mirror symmetry in the hopping rates via $J_B=J_A+\delta$ (cf.~the discussion below Eq.~\eqref{eq:Hxx}).
        We also consider the addition of both equal detunings and opposite detunings to the second site of each chain: $\hat{H}_\Delta = (\delta/2)(\hat{\sigma}_{A,2}^z \pm \hat{\sigma}_{B,2}^z)$.
        The purity of the numerically computed steady state $\hat{\rho}_{\rm ss}=\hat{\rho}(t\to\infty)$ is shown in each case vs.~the perturbation strength $\delta$. The steady state is only pure when $\delta=0$.
        We use unperturbed system parameters $\Omega=0.5\gamma$, $\Delta=0.1\gamma$, and $J=0.25\gamma$.
    }
     \label{fig:perturbation-purity}
 \end{figure}

We now introduce a key result of this work:  the boundary-driven double spin chain in Eq.~\eqref{eq:qme} has a pure steady state for {\it arbitrary} $N$, drive strength $\Omega$ and hoppings $J_j$.  
Even though the $1{+1}$-qubit system has generically a unique pure steady state, a priori there is no reason to expect that this will also be true when $N > 1$.  Indeed, Fig.~\ref{fig:perturbation-purity} shows that for a generalized version of our $2+2$ qubit model, the steady state will be impure if the two hopping amplitudes differ, or if we detune the second pair of qubits from the drive.  

We find surprisingly that these two conditions (mirror symmetry of hoppings, no detunings of additional qubits) are enough to guarantee a pure steady state for {\it arbitrary} $N$, and for arbitrary choices of the parameters in Eq.~\eqref{eq:qme}.  In the infinite-drive limit, the steady state has a simple translationally-invariant dimerized form that can be understood from a general replication argument that we present below (and that applies to many other setups
\cite{pocklington_Stabilizing_2022,zippilli_Entanglement_2013,zippilli_Steadystate_2015,angeletti_Dissipative_2023}).  For finite drives $\Omega$, the steady state has a far more complicated form that is neither dimerized nor translationaly invariant.  Our exact analytic expression nonetheless provides a simple picture for the state:  it is a condensate of paired ``hole" excitations, where holes correspond to cross-chain dimers that are in the vacuum $|(00)_j\rangle$ state. 

%%%%%%%%%%%%%%%%%%%%%%%%%%%%%%%%%%%%%%%
%%%%%%%%%%%%%%%%%%%%%%%%%%%%%%%%%%%%%%%

%%%%%%%%%%%%%%%%%%%%%%%%%%%%%%%%%%%%%%%%%%%%%%%%%%%%%%%%%%%%%%%%%%%%%%

\subsection{\texorpdfstring{$\Omega\to\infty$}{\textOmega~=~infinity}  limit: generic entanglement replication via {\it XX} couplings}
\label{sec:replication}

The form of this pure state of Eq.~\eqref{eq:qme} becomes extremely simple in the limit of strong driving $\Omega$:  
\begin{align}
    | \psi \rangle \rightarrow |\psi_\infty\rangle &= |S_{1}T_{2}S_{3}T_{4}S_{5}T_{6}\cdots(S/T)_{N}\rangle.
    \label{eq:steady-state-strong-drive}
\end{align}
This is a highly entangled state of the two chains, that factors into a product of Bell pairs on each cross-chain dimer $j$.  The phase of these pairs alternates from $|S\rangle$ or $|T\rangle$ as one moves down the chain as indicated.  Up to this local phase variation, the state is translationally invariant.  
This result is in fact the consequence of a much more general ``entanglement replication'' phenomenon associated XX couplings and definite-parity dimer states; we explain this in what follows.  
This mechanism also explains and unifies the replication phenomena seen in several previous works  \cite{pocklington_Stabilizing_2022,zippilli_Entanglement_2013,zippilli_Steadystate_2015,angeletti_Dissipative_2023}.  We note that the general nature of the replication mechanism we present here was not discussed in earlier works.

\begin{figure}[t!]
     \centering
    \includegraphics[width=0.98\columnwidth]{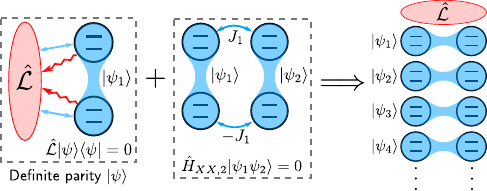}
     \caption{
        \textbf{Entanglement replication via passive exchange couplings.}
        Suppose there is some dissipative dynamics $\hat{\mathcal{L}}$ that stabilizes a two qubit entangled state $|\psi\rangle$ on the first pair of qubits in two exchange-coupled chains. 
        If that state has definite parity, then the product state $|\psi_1\psi_2\rangle$ is a zero energy eigenstate of passive exchange couplings $\hat{H}_{\rm XX}$ (cf. Eq.~\eqref{eq:rep-HXX2}). 
        Thus the product state is a steady state of the stabilization dynamics and exchange couplings.
        Therefore, any number of qubit pairs may be added via exchange couplings, resulting in the tensor product steady state $|\psi_1\psi_2\cdots\psi_N\rangle$, replicating the entanglement on site 1 down a pair of arbitrarily long chains. 
    }
     \label{fig:replication}
 \end{figure}

Imagine, as in Fig.~\ref{fig:replication}, that we have dissipative dynamics $\hat{\mathcal{L}}$ acting on a 2-qubit system that stabilizes an arbitrary state $|\psi \rangle$ with fixed excitation number parity, i.e. $|\psi \rangle = a |00\rangle + b |11\rangle$ or $|\psi \rangle = a |01\rangle + b |10\rangle$. The actual method for stabilization is unimportant, only the state matters. For now, we will focus on even parity states for concreteness, but the analysis is identical for odd parity ones.  Note for our specific system, for the $N=1$ case and $\Omega \rightarrow \infty$, the stabilized state has a definite parity, hence the following arguments apply.  

Next, imagine passively coupling a second pair of qubits to the first with an XX Hamiltonian, c.f. Eq.~\eqref{eq:Hxx}.  
For convenience, we use a different gauge choice for the $B$ chain (i.e. $|1\rangle \rightarrow -|1\rangle$ for the $B$ qubits), such that the XX couplings now have opposite signs in the $B$ chain versus the $A$ chain.
Specifically, if $\hat{\mathcal{L}}$ stabilizes $|\psi_1 \rangle = a|(00)_{1}\rangle + b|(11)_{1}\rangle $, then we are interested in the Hamiltonian
\begin{align}
    \hat H_{\mathrm{XX},2} &= J \sum_{s = A,B} (-1)^s \left[ \hat{\sigma}^x_{s,1}\hat{\sigma}^x_{s,2} + \hat{\sigma}^y_{s,1}\hat{\sigma}^y_{s,2} \right].
    \label{eq:rep-HXX2}
\end{align}
This Hamiltonian now has an extremely convenient feature:  no matter what the parameters $a,b$, the ``replicated", dimerized state $|\psi \rangle = |\psi_1\psi_2\rangle$ is a zero eigenstate, 
as can be confirmed by a simple direct computation
\begin{align}
    \hat H_{\mathrm{XX},2} |\psi_1 \psi_2 \rangle &= 0.
\end{align}
This tells us that $|\psi \rangle = |\psi \rangle_1 \otimes |\psi \rangle_2$ is a steady state of the 4-qubit dynamics defined by 
\begin{align}
    \partial_t \hat \rho &= \left( \hat{\mathcal{L}} \otimes \mathds{1} \right) \hat \rho - i[\hat H_{\mathrm{XX},2}, \hat \rho].
\end{align}
Hence, without changing the dissipative stabilization mechanism at all, we can use the passive Hamiltonian interaction to propagate entanglement to a second qubit pair. Even more strikingly, we can now add a third pair of qubits (again via mirrored XX couplings).  The same arguments tell us that the steady state will be a replicated entangled state, i.e. a product of cross-chain dimers, where each entangled dimer is in the state $| \psi \rangle$.  Iterating this argument, we can show that for arbitrary $N$, if we let
\begin{align}
    \hat H_{\mathrm{XX},N} &= \sum_{i = 1}^{N - 1} \sum_{s = A,B}  J_i (-1)^s \left[  \hat{\sigma}^x_{s,i}\hat{\sigma}^x_{s,i + 1} + \hat{\sigma}^y_{s,i}\hat{\sigma}^y_{s,i + 1} \right], \label{eqn:XX_Ham} \\
    |\psi \rangle &= \bigotimes_{i = 1}^N \bigg[ a|(00)_{i}\rangle + b|(11)_{i}\rangle \bigg], \label{eqn:replicated_state}
\end{align}
then $|\psi \rangle$ is still a steady state of the dynamics
\begin{align}
    \left( \hat{\mathcal{L}} \otimes \mathds{1}^{\otimes N-1} \right) |\psi \rangle \langle \psi | - i \left[ \hat H_{\mathrm{XX},N}, |\psi \rangle \langle \psi | \right] = 0.
\end{align}
The analysis follows in the exact same manner if one uses odd parity states instead of even parity ones. 
Elsewhere in this manuscript, we work in a gauge with uniform coupling signs (so that there is no factor of $(-1)^s$ in Eq.~\eqref{eq:Hxx}, contrast with Eq.~\eqref{eqn:XX_Ham}). 
The replication here goes through exactly the same, where we can make a local sign flip on every other dimer, moving the relative phase onto the $b$ coefficient in Eq.~\eqref{eqn:replicated_state},  $b \to (-1)^i b$.
Therefore, this replication argument gives a general proof that Eq.~\eqref{eq:steady-state-strong-drive} is a pure steady state of Eq.~\eqref{eq:qme} in the infinite driving limit (as in this limit, the $N=1$ problem has a definite (odd) parity pure steady state $|S_1\rangle$)
\footnote{The argument here does not prove uniqueness of the replicated steady state.  In many cases, including that studied here, the lack of additional symmetries precludes additional steady states.}.
The change of gauge thus explains why Eq.~\eqref{eq:steady-state-strong-drive} is a product of staggered $|S\rangle$ and $|T\rangle$ instead of uniform $|S\rangle$ states.

%%%%%%%%%%%%%%%%%%%%%%%%%%%%%%%%%%%%%%%
%%%%%%%%%%%%%%%%%%%%%%%%%%%%%%%%%%%%%%%

The fact that the Hamiltonian perfectly replicates fixed parity states can be understood intuitively from the fact that it commutes with $\hat S^2_{s}$ the total spin operator and $\hat S^z_{s} = \hat \sigma_{s,1}^z + \hat \sigma_{s,2}^z$ the collective $Z$ operator of each chain $s=A,B$. More details are in App.~\ref{app:XXintuition}.
More generally, one can demonstrate that given any {\it arbitrary} two-qubit entangled state, Heisenberg couplings can be used to perfectly replicate this state down the chain, alleviating the parity constraint. Details are in App.~\ref{app:heisreplication}. Moreover, for both the Heisenberg coupling or XX couplings, significantly more complex geometries than chains can be used, generalizing \cite{angeletti_Dissipative_2023}. For more details, see App.~\ref{app:replication-trees}.

%%%%%%%%%%%%%%%%%%%%%%%%%%%%%%%%%%%%%%%%%%%%%%%%%%%%%%%%%%%%%%%%%%%%%%

\subsection{\texorpdfstring{$\Omega < \infty$}{\textOmega~\textless~infinity}: pure steady-state condensate of paired holes}

We now turn to the more general (and experimentally relevant) case of a non-infinite drive amplitude $\Omega$.
For finite strength driving, the steady state for $N=1$, Eq.~\eqref{eq:steady-state-2qb}, no longer has definite parity.  As such, the replication argument of the previous section does not apply when we now consider larger systems, and there are no general arguments that would guarantee the existence of a pure steady state for $N \geq 2$.  Remarkably, we find that for arbitrary parameters, Eq.~\eqref{eq:qme} has a pure steady state, albeit one that is far more complicated than the dimerized, translationally invariant state in Eq.~\eqref{eq:steady-state-strong-drive}.  As we now show, this state can be exactly understood as a condensate of paired ``holes".  Here, a ``hole pair'' excitation is defined as starting with the ``filled" state in Eq.~\eqref{eq:steady-state-strong-drive}, and then replacing the state of two adjacent dimers with the vacuum state, i.e.~$|(S)_j\rangle |(T)_{j+1} \rangle \to |(00)_j\rangle |(00)_{j+1}\rangle$.  

Certain features of our state can be understood intuitively.  
For finite $\Omega$, it is reasonable to expect the presence of holes, i.e. fewer qubit excitations than in the infinite driving limit.
Further, these holes should be delocalized throughout the chain in order to have an eigenstate of the kinetic energy term $\hat{H}_{\rm XX}$.
This motivates looking for a pure steady state having delocalized hole excitations (with the density of holes scaling inversely with drive amplitude).  
More unexpected is our finding that in the steady state, these holes must be paired on adjacent sites.

To present our solution, it is convenient to map the double chain system to a 1D ``dimer chain'', where each site of the new 1D chain has local Hilbert space dimension 4, and corresponds to a dimer of the original system. 
With this mapping, we introduce two flavors of ``particle'' $|\bullet_j\rangle$ and $|{\sqr}_j\rangle$ and ``hole'' $|\circ_j\rangle$ states via
\begin{align}
    |\circ_j\rangle &\equiv |(00)_j\rangle, \label{eq:hole}\\ 
    |\bullet_j\rangle &\equiv \frac{1}{\sqrt{2}}\hat{\tau}_{j}^\dagger|\circ_j\rangle =
    \begin{cases}
        |S_j\rangle & j~{\rm odd} \\ 
        |T_j\rangle & j~{\rm even}
    \end{cases},
    \label{eq:bullet-particle} \\
    |{\sqr}_j\rangle &\equiv \frac{1}{\sqrt{2}}\hat{\lambda}_{j}^\dagger|\circ_j\rangle =
    \begin{cases}
        |T_j\rangle & j~{\rm odd} \\ 
        |S_j\rangle & j~{\rm even}
    \end{cases},
    \label{eq:sqr-particle}
\end{align}
where we implicitly define the dimer ladder operators $\hat{\tau}_{j}^\dagger,\hat{\tau}_{j}$ that create and destroy the dimer particles $|\bullet_j\rangle=|(S/T)_j\rangle$ (for $j$  odd/even) when acting on $|\circ_j\rangle$ or $|\bullet_j\rangle$, respectively.
Similarly $\hat{\lambda}_j^\dagger,\hat{\lambda}_j$ create and destroy the particles $|{\sqr}_j\rangle=|(T/S)_j\rangle$ (for $j$ odd/even) when acting on $|\circ_j\rangle$ or $|{\sqr}_j\rangle$, respectively.
Note that there is a hard-core constraint that prevents a $|{\bullet}\rangle$ and a $|{\sqr}\rangle$ from simultaneously occupying a site.
We can neglect the remaining basis state for each dimer for now, as this state does not appear in the pure steady state of interest (see App.~\ref{app:dimer-rep} for more details).  
With our new representation, the filled state Eq.~\eqref{eq:steady-state-strong-drive} is thus $|\psi_\infty\rangle=|\bullet\bullet\bullet\cdots\rangle$.

Using the dimer particle representation defined in Eq.~\eqref{eq:bullet-particle}, we introduce an operator that creates a delocalized hole-pair:
\begin{align}
    \hat{Q} = \frac{1}{2\Jbar\sqrt{N}}\sum_{j=1}^{N-1} J_{j} (-1)^{j}  \hat{\tau}_{j}\hat{\tau}_{j+1}.
    \label{eq:Qpair}
\end{align}
Here 
\begin{align}
    \Jbar = \sqrt{\frac{1}{N-1}\sum_j J_j^2}
    \label{eq:Jbar}
\end{align}
is the RMS hopping rate. 
Acting on the reference ``filled" state $|\psi_\infty\rangle$, the operator $\hat{Q}$ creates a superposition state where each term corresponds to a pair of adjacent holes at a different location
\footnote{For convenience, we normalize the operator by $1/\Jbar\sqrt{N}$ so it produces an approximately normalized state when acting on $|\psi_\infty\rangle$ and other nearly-filled states (${||}\hat{Q}|\psi_\infty\rangle{||}^2\approx {||}\hat{Q}^2|\psi_\infty\rangle{||}^2 \approx 1$).}.  

At a heuristic level, the phases in $\hat{Q}$ will give each hole pair a net momentum, allowing them to become zero-energy eigenstates of the ``kinetic energy" $\hat{H}_{\rm XX}$.  More formally, 
we show in App.~\ref{app:hole-pairing} that $\hat{Q}$ commutes with $\hat{H}_{\rm XX}$.  Its action on an $\hat{H}_{\rm XX}$ eigenstate will thus produce another eigenstate with the same energy.  In particular:
\begin{align}
    \hat{H}_{\rm XX}(\hat{Q}|\psi_\infty\rangle) = \hat{Q} \hat{H}_{\rm XX}|\psi_\infty\rangle = 0.
\end{align}
An entire tower of zero-energy $\hat{H}_{\rm XX}$ eigenstates can thus be generated by repeated application of $\hat{Q}$ to $|\psi_\infty\rangle$, each state having 
an increasing number of hole pairs.  The maximal-hole state in this tower corresponds to the empty state (if $N$ is even), or a state with a single delocalized particle $|\bullet_j\rangle$ (if $N$ is odd).

%%%%%%%%%%%%%%%%%%%%%%%%%%%%%%%%%%%%%%%
 \begin{figure}[t!]
     \centering
    \includegraphics[width=0.98\columnwidth]{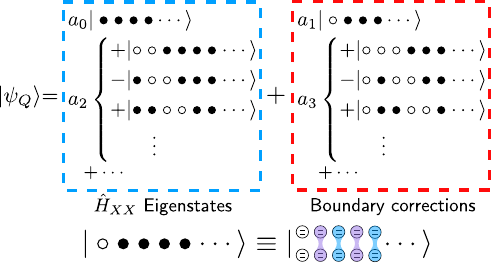}
     \caption{
        \textbf{First few terms of the steady state.}
        The expansion of the steady state in the number of holes added to $|\psi_\infty\rangle$ (cf. Eq.~\eqref{eq:steady-state-strong-drive}). 
        The expansion coefficients $a_n\sim 1/\drive^n$ are the weights of the $n$th-hole components of the state; they can be found analytically from Eq.~\eqref{eq:steady-state}.
        Here we take uniform $J_j = \Jbar$.
        For nonuniform $J_j$ the components of each $a_n$ term are weighted by factors of $J_j/\Jbar$.
    }
     \label{fig:steady-state}
 \end{figure} 
%%%%%%%%%%%%%%%%%%%%%%%%%%%%%%%%%%%%%%%

Recall that in general, to obtain a pure steady state we require a state that is both an eigenstate of the Hamiltonian, and annihilated by relevant dissipators.  The family of states $\hat{Q}^m |\psi_\infty\rangle$ provide us with a large class of states that are compatible with the Hamiltonian and connected to the pure steady state in the infinite-drive limit.  One might expect that they can be used to construct the steady state for the finite-drive amplitude system.  We find that apart from a boundary correction, this is indeed the case.  
As we rigorously show in App.~\ref{app:steady-state-proof}, for any set of parameters, Eq.~\eqref{eq:qme} has a pure steady state $|\psi_Q\rangle$ given by the ``pair condensate" state
\begin{align}
    |\psi_Q\rangle &= \left(1 + \frac{\Gamma}{\Omega}\hat{\tau}_{1}\right) \exp\left[\frac{\sqrt{N}}{\drive^{2}}\hat{Q}\right]|\psi_\infty\rangle
    \label{eq:steady-state}
\end{align}
where $\hat{\tau}_1$ is the dimer lowering operator that removes the particle on site 1 (cf. Eq.~\eqref{eq:bullet-particle}) and the reference state $|\psi_\infty\rangle$ is given by Eq.~\eqref{eq:steady-state-strong-drive}.
The dimensionless drive strength $\drive$ appearing in the exponential is
\begin{align}
    \drive &\equiv \frac{\Omega}{\sqrt{\Gamma \Jbar}},
    \label{eq:eff-drive}
\end{align}
with $\Gamma$ given by Eq.~\eqref{eq:Gamma} and $\Jbar$ by Eq.~\eqref{eq:Jbar}.
The first few terms of $|\psi_Q\rangle$ are shown in Fig.~\ref{fig:steady-state}.
Up to overall normalization, the coefficients $a_j$ can be read off from Eq.~\eqref{eq:steady-state}, e.g., for a uniform chain ($J_j=\Jbar$) $a_0 = 1$, $a_1 = \Gamma/\Omega = \sqrt{\Gamma/\Jbar}/\drive$, $a_2 = 1/\drive^2$, etc; this is a power series in $1/\drive$, with $a_j\sim 1/\drive^j$.
As we will discuss in more detail, this exact solution also immediately lets us understand the NESS of the non-trivial single chain system in Fig.~\ref{fig:intro}(b).

Finally, we note that there is an equivalent construction of $|\psi_Q\rangle$ that is recursive in the length $N$ of the chains.
The recursive construction enables the efficient numerical evaluation of expectation values and correlation functions, e.g. particle density $\langle\hat{n}_j\rangle = \frac{1}{2}\langle\hat{\tau}_j^\dagger\hat{\tau}_j\rangle$.
Details are provided in App.~\ref{app:recursion-correlation}.

%%%%%%%%%%%%%%%%%%%%%%%%%%%%%%%%%%%%%%%%%%%%%%%%%%%%%%%%%%%%%%%%%%%%%%
%%%%%%%%%%%%%%%%%%%%%%%%%%%%%%%%%%%%%%%%%%%%%%%%%%%%%%%%%%%%%%%%%%%%%%

\section{Real-space hole pairing and  consequences for the steady state}
\label{sec:hole-pairing}

\subsection{Hole pairing as a kinetic constraint}

%%%%%%%%%%%%%%%%%%%%%%%%%%%%%%%%%%%%%%%
 \begin{figure}[t!]
     \centering
    \includegraphics[width=0.98\columnwidth]{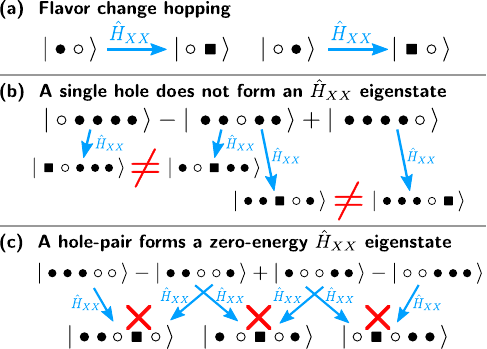}
     \caption{
        \textbf{Flavor-change hopping forces hole-pairing in zero-energy eigenstates.}
        \textbf{(a)} $\hat{H}_{\rm XX}$ (cf. Eq.~\eqref{eq:Hxx}) swaps a particle and a hole in the dimer chain and changes the particle flavor in the process.
        \textbf{(b)} Due to the flavor change in hopping, a single delocalized hole is not an eigenstate of the chain because the final states of a hole hopping to the left and hole hopping to the right are distinguishable: the $|{\sqr}\rangle$ either ends up to the right of the hole or to the left of the hole, respectively.
        \textbf{(c)} Adjacent paired holes can form zero-energy eigenstates via destructive interference because the $|{\sqr}\rangle$ always ends up sandwiched between two holes.
    }
     \label{fig:hole-pairing}
 \end{figure} 
%%%%%%%%%%%%%%%%%%%%%%%%%%%%%%%%%%%%%%%

The hole-pairing in $|\psi_Q\rangle$ is surprising at first glance, as there is no attractive interaction or other explicit pairing mechanism in our model.  
Hole pairing turns out to be a consequence of two facts: (i) $\hat{H}_{\rm XX}$ causes $|{\bullet}\rangle$ and $|{\sqr}\rangle$ particles to change flavor when they hop, and (ii) the hard-core constraint forbids a $|{\bullet}\rangle$ and an $|{\sqr}\rangle$ to simultaneously occupy a site.
As shown in Fig.~\ref{fig:hole-pairing}(a), with details in App.~\ref{app:dimer-rep}, a particle can swap positions with a hole and change flavor in the process.
Two adjacent particles of the same flavor cannot hop due to the hard-core constraint, hence $\hat{H}_{\rm XX}|\!\bullet\bullet\rangle = 0$.
If we now try to form zero kinetic energy states (i.e.,~zero-energy eigenstates of $\hat{H}_{\rm XX}$) by delocalizing hole excitations, we find that they  \emph{must} be paired on adjacent sites.
Delocalizing a single hole to get a zero-energy eigenstate fails due to the flavor-changing hopping, see Fig.~\ref{fig:hole-pairing}(b), 
but delocalizing a pair is successful (see Fig.~\ref{fig:hole-pairing}(c)).

The creation of zero kinetic energy eigenstates of $\hat{H}_{\rm XX}$ via the hole-pairing operator $\hat{Q}$ (cf. Eq.~\eqref{eq:Qpair}) is somewhat analogous to $\eta$-pairing found in Fermi-Hubbard lattices \cite{yang_eta_1989,zhang_Pseudospin_1990}.
In both cases, the pairing operator generates exact Hamiltonian eigenstates with zero kinetic energy by delocalizing a pair of excitations throughout the system.
A key distinction however is that in $\eta$-pairing, each paired excitation is spatially local, i.e.~a pair of fermions on one lattice site, whereas in $Q$-pairing, each hole-pair occupies adjacent lattice sites.
We also note that the algebraic structure of $\eta$-pairing ($\hat{\eta}, \hat{\eta}^\dagger$ form a closed representation of SU(2)) is not found in $Q$-pairing: $\hat{Q}$ and $\hat{Q}^\dagger$ do not form a closed SU(2) group.

While not a true symmetry of the Hamiltonian, the hole-pairing operator has a special relation to the Hamiltonian of the double-chain system, via a structure that was introduced in the context of many-body scar states.  Specifically, it constitutes a restricted spectrum generating algebra (RSGA) of double-chain Hamiltonian when acting on $|\psi_\infty\rangle$ \cite{moudgalya_etapairing_2020}.  Furthermore, certain terms can be added to the XX chain to make the model nonintegrable while preserving this RSGA structure, thus guaranteeing that $\hat Q^n|\psi_\infty\rangle$ remain exact eigenstates. In App.~\ref{app:rsga-scar} we discuss how this makes the corresponding hole-pairing states true many-body scar states in a nonintegrable ladder system (a model related to that studied in Ref.~\cite{znidaric_Coexistence_2013}).

Having understood the route to hole pairing in our model, we can also postulate other Hamiltonian models where this will occur.  For example, 
a 1D Fermi-Hubbard chain with a spin orbit interaction can exhibit effective flavor-changing hopping. For strong interactions, it can thus also exhibit hole-pairing in a subset of its eigenstates (see App.~\ref{app:hole-pairing}).
The fermionic analog of the hole-pairing operator $\hat{Q}$ has the same properties, generating eigenstates of the 1D chain when acting on a filled state.
These hole-paired eigenstates may be accessible in e.g. the ultracold atoms platform proposed in Ref.~\cite{mamaev_Quantum_2019}.

\subsection{Density correlations due to hole pairing}

\begin{figure}[t!]
     \centering
    \includegraphics[width=0.98\columnwidth]{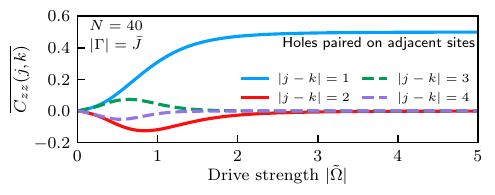}
     \caption{
        \textbf{Density correlations on adjacent sites are a direct observable consequences of hole pairing.}
        The hole density correlation function $C_{zz}(j,k)$ (cf.~Eq.~\eqref{eq:hole-density-correlation-single}) is averaged over averaged over the entire $N=40$ chain, while holding $|k-j|$ fixed. The averaged $\overline{C_{zz}(j,k)}$ is plotted vs.~drive strength $|\drive|$.
        The nearest neighbor ($|k-j|=1$) correlations saturate to $0.5$ with increasing $|\drive|$, indicating that as the filled state $|\psi_\infty\rangle$ is approached, deviations from $|\psi_\infty\rangle$ in the bulk are due to holes paired on adjacent sites.
        For this plot we hold $|\Gamma| = \Jbar$ fixed.
    }
     \label{fig:hole-pairing-correlations}
 \end{figure}

The structured hole pairing 
in  Eq.~\eqref{eq:steady-state} immediately gives rise to spatial density correlations, something that is most apparent when the hole density is low, i.e.~$|\drive|\gtrsim 2$.   This correlation is directly observable in the single-chain system of Fig.~\ref{fig:intro}(a) as 
$Z$ magnetization correlations between adjacent sites, $\langle\hat{\sigma}^z_{A,j}\hat{\sigma}^z_{A,j+1}\rangle \neq 0$.
The correspondence between magnetization and hole density follows from the fact that the dimer holes $|\circ\rangle$ are polarized but the particles $|\bullet\rangle$ are depolarized (cf.~Eqs.~\eqref{eq:hole} and \eqref{eq:bullet-particle}):
\begin{align}
    \langle\bullet_j|\hat{\sigma}^z_{A,j}|\bullet_j\rangle=0;\qquad\langle\circ_j|\hat{\sigma}^z_{A,j}|\circ_j\rangle=-1.
\end{align}
Thus, $(-\hat{\sigma}^z_{A,j})$ acts as a local hole number operator when acting on the steady state.
We define the $z$-magnetization correlation function for the $A$ chain as
\begin{align}
    C_{zz}(j,k) = \frac{\langle \hat{\sigma}^z_{A,j}\hat{\sigma}^z_{A,k}\rangle - \langle \hat{\sigma}^z_{A,j}\rangle\langle\hat{\sigma}^z_{A,k}\rangle}{\sqrt{\big(\langle\hat{\sigma}_{A,j}^z\rangle+\langle\hat{\sigma}_{A,j}^z\rangle^2\big)\big(\langle\hat{\sigma}_{A,k}^z\rangle+\langle\hat{\sigma}_{A,k}^z\rangle^2\big)}},
    \label{eq:hole-density-correlation-single}
\end{align}
normalized such that $C _{zz}(j,j)=1$.
See App.~\ref{app:hole-correlation} for a discussion of the nonstandard normalization.
In Fig.~\ref{fig:hole-pairing-correlations}, we show $C_{zz}(j,k)$ for fixed distance $|k-j|$ and averaged over the whole chain.
For strong driving, the average correlations between adjacent sites saturates to $\overline{C_{zz}(j,j\pm1)}\to 0.5$ because in this regime, a hole on site $j$ is always paired with a hole on either $j\pm1$, but is unlikely to be correlated with any other site.  In contrast, there are no appreciable correlations in this regime for larger distances.
We note that the correlation functions $C_{zz}(j,k)$ can be measured in either the double-chain system or the single-chain system. In either case, the saturation of $C_{zz}(j,j\pm 1)\approx 0.5$ for $\tilde{\Omega}>1$ is a clear indication of hole pairing.

%%%%%%%%%%%%%%%%%%%%%%%%%%%%%%%%%%%%%%%
%%%%%%%%%%%%%%%%%%%%%%%%%%%%%%%%%%%%%%%

%%%%%%%%%%%%%%%%%%%%%%%%%%%%%%%%%%%%%%%%%%%%%%%%%%%%%%%%%%%%%%%%%%%%%%

\subsection{Universal density scaling}
\label{subsec:univ-scaling}

There are two dimensionless parameters in our model:  the dimensionless drive amplitude $\drive$ (c.f.~Eq.~\eqref{eq:eff-drive}) and the ratio $\zeta \equiv \sqrt{\Gamma/\Jbar}$.  One might naturally expect that bulk steady state properties would depend on both these parameters.  
However, the exact result Eq.~\eqref{eq:steady-state} shows that this is not the case.  We can write this as
\begin{align}
    |\psi_Q\rangle &= \left(1 + \frac{\zeta}{\drive} \hat{\tau}_1\right) \exp\left[ \frac{\sqrt{N}}{\drive^2}\hat{Q} \right]|\psi_\infty\rangle, 
    \label{eq:ss-rewrite}
\end{align}
which suggests that the excitation density in the bulk is controlled {\it only} by $\drive$.  
We now show explicitly that the excitation density (i.e.~$Z$ magnetization density) is indeed intensive,  and scales universally with the single parameter $|\drive|$ in the regime $|\drive|\gtrsim 2$.

Consider first the limit where  $|\zeta|\to 0$ while $|\drive|$ remains fixed, such that we can ignore the boundary term in Eq.~\eqref{eq:ss-rewrite}.  We thus have $|\psi_Q\rangle = e^{\alpha \hat{Q}}|\psi_\infty\rangle$, where $\alpha=\sqrt{N}/|\drive|^2$.
This expression mimics a bosonic coherent state $|\alpha\rangle = e^{\alpha \hat{a}^\dagger}|0\rangle$, where  hole pairs are the bosons, $|\psi_\infty\rangle$ is the hole-pair vacuum and $\hat{Q}$ is the hole-pair creation operator.
As we show in App.~\ref{app:univ-state-norms}, this analogy to bosonic coherent states can be made precise when $N \gg 1$ and the hole density $\bar{m}$ satisfies
\begin{align}
    \bar{m} &\equiv \frac{1}{N}\sum_j \langle-\hat{\sigma}^z_{A,j}\rangle \ll 1.
    \label{eq:hole-density}
\end{align}
In this low hole density regime, we may neglect the hard-core constraint that prevents two hole-pairs from occupying the same sites, and can thus accurately estimate the total hole number as 
\begin{align}
    M = 2|\alpha|^2 = \frac{2N}{|\drive|^4}
    \equiv N \bar{m}.
    \label{eq:universal-scaling}
\end{align}
We thus find for large drives an intensive scaling of hole density.  
Note that this result immediately implies that for an arbitrarily long single chain system, 
one only needs $|\tilde{\Omega}| \gtrsim 2$ to approach the infinite temperature state.

\begin{figure}[t!]
     \centering
    \includegraphics[width=0.98\columnwidth]{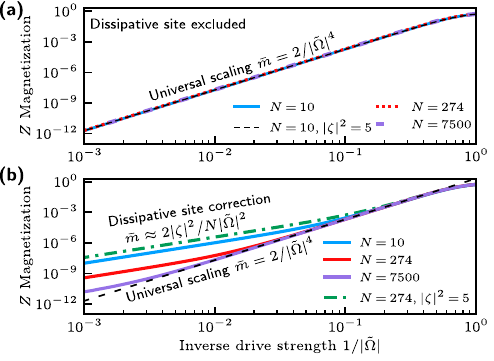}
     \caption{
        \textbf{Universal scaling of steady state $Z$ magnetization (hole density) }
        The $Z$ magnetization / hole density $\bar{m}$ (cf. Eq.~\eqref{eq:hole-density}) is shown for chains of widely varying length as a function of inverse effective drive strength $1/|\drive|$.
        \textbf{(a)} Here, we plot the hole density, excluding the dissipative site, which shows universal scaling $\bar{m} = 2/|\drive|^4$ for $1/|\drive|\gtrsim 1$.
        \textbf{(b)} Here, we include the dissipative site and show that $\bar{m}$ now has deviations $\sim 1/N|\drive|^2$ appearing when $|\drive|^2\approx N/|\zeta|^2$.
        For both plots, $|\zeta|^2\equiv |\Gamma/\Jbar| = 0.05$ except where indicated otherwise, and we vary only $\Omega$.
    }
     \label{fig:universal-density}
 \end{figure}

When $|\zeta|>0$, the coherent state analogy is still valid, but there is a boundary correction to Eq.~\eqref{eq:universal-scaling}. Because site 1 directly sees the dissipation, it can be occupied by an isolated hole and thus its occupation depends not only on $|\drive|$ but also $|\zeta|$.
The hole density $\bar{m}$ thus obtains a $\mathcal{O}(1/N)$ correction to the intensive universal scaling
\begin{align}
    \bar{m} = \frac{2}{|\drive|^4} + \frac{2|\zeta|^2}{N(|\drive|^2+2|\zeta|^2)}.
\end{align}
There is thus a transition from universal $\sim|\drive|^{-4}$ scaling to an $N$-dependent $\sim|\drive|^{-2}$ scaling at $|\drive|^2\approx N/|\zeta|^2$, or equivalently $\Omega/\Jbar \approx \sqrt{N}$.
The universal scaling behavior is exact for any $|\zeta|$ when the dissipative site density is excluded, as is shown in Fig.~\ref{fig:universal-density}(a), and the dissipative site corrections for $|\zeta|>0$ are shown in Fig.~\ref{fig:universal-density}(b).

%%%%%%%%%%%%%%%%%%%%%%%%%%%%%%%%%%%%%%%
%%%%%%%%%%%%%%%%%%%%%%%%%%%%%%%%%%%%%%%

%%%%%%%%%%%%%%%%%%%%%%%%%%%%%%%%%%%%%%%%%%%%%%%%%%%%%%%%%%%%%%%%%%%%%%

\subsection{Single particle ``charge density waves''}

For $J_j = J$, the spin chain systems in Fig.~\ref{fig:intro} are translationally invariant except for the boundary $j=1$ site.  For long chains, one might thus expect that the steady state density is also translationally invariant, except for edge effects near $j=1$.  Surprisingly, this expected translational invariance is strongly broken in the steady state for weak drives $\drive$.   
As a consequence of the hole pair condensate of Eq.~\eqref{eq:steady-state}, there is a regime where the steady state corresponds to a single excitation that is localized on either the even-$j$ or odd-$j$ sublattice, i.e.~a kind of single particle charge density wave (CDW).  

As we will show, for weak drives, the double spin chain system's steady state (for uniform $J_j=\Jbar$) is given by the CDW form:
\begin{align}
    |\Psi_{\rm cdw}\rangle = \frac{1}{\sqrt{\lceil N/2 \rceil}} \sum_{j}(-1)^j|\bullet_{N-2j}\rangle,
    \label{eq:dimer-cdw}
\end{align}
where $\lceil X \rceil$ is the ceiling function, and $|\bullet_j\rangle$ is given by Eq.~\eqref{eq:bullet-particle}.
This describes a single particle that is delocalized either across all odd-$j$ sites in odd-length chains or across all even-$j$ sites in even-length chains.
For disordered $J_j$, each component $|\bullet_j\rangle$ is weighted by an additional factor $(\cdots J_{j-4}J_{j-2})(J_{j+1}J_{j+3}\cdots)/\Jbar^{\lceil (N-1)/2\rceil}$ (and the requisite correction to the normalization).

For the single dissipative spin chain, Fig.~\ref{fig:intro}(b), the non-equilibrium steady state $\hat{\rho}_{A,{\rm cdw}}= {\rm tr}_B |\Psi_{\rm cdw}\rangle\langle\Psi_{\rm cdw}|$ is an equal mixture of vacuum and the single-particle CDW:
\begin{align}
    \hat{\rho}_{A,{\rm cdw}} &= \frac{1}{2}\Big[|0\rangle\langle0| + |\Phi_{\rm cdw}\rangle\langle\Phi_{\rm cdw}|\Big], 
    \label{eq:cdw-density-matrix}\\
    |\Phi_{\rm cdw}\rangle &= \frac{1}{\sqrt{\lceil N/2 \rceil}} \sum_{j}(-1)^j \hat{\sigma}_{A,N-2j}^+|0\rangle.
\end{align}
Despite not being a pure state, the single chain CDW is a very low entropy state, $S(\hat{\rho}_{A,{\rm cdw}}) = \ln 2$, for which any particle density necessarily arises from the single coherently delocalized excitation.
Here, $|\Phi_{\rm cdw}\rangle$ is the single chain equivalent of $|\Psi_{\rm cdw}\rangle$ given above, and its components have the same weighting factors when the $J_j$ are disordered.

%%%%%%%%%%%%%%%%%%%%%%%%%%%%%%%%%%%%%%%
 \begin{figure}[t]
    \centering
    \includegraphics[width=0.98\columnwidth]{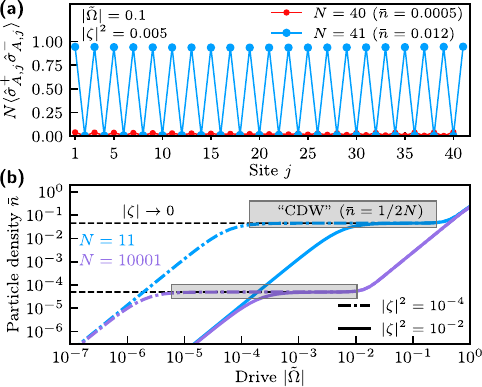}
     \caption{
        \textbf{The emergence of single particle ``CDWs'' in finite length spin chains.}
        \textbf{(a)} Local particle density $\langle\hat{n}_j\rangle$ is plotted for each site of 40 and 41 site chains, scaled by length $N$.
        For the given parameters, $\drive=0.1$ and $|\zeta| = 0.005$, and taking uniform $J_j$,  the odd-length chain has a single particle delocalized across all odd sites, hence the average particle density is $\bar{n} = 1/2N \approx 0.012$.
        The even length chains do not have charge density waves, and have much smaller average density $\bar{n} = N|\drive|^4/8 \approx 0.0005$.
        The modulation of local density across the chain is due to the highly-correlated two-particle state.
        \textbf{(b)}
        Particle density $\bar{n}$ is plotted vs. drive strength $|\drive|$ for two odd-length chains ($N=11$ and $N=10001$) in the odd-length CDW driving strength regime (cf. Eq.~\eqref{eq:cdw-regime-odd}) for two different $|\zeta|^2=10^{-4}$ and $|\zeta|^2=10^{-2}$.
        In the limit $|\zeta|\to0$ (dashed curves), the single particle CDW persists for any arbitrarily small $|\drive|>0$.
        In both plots, the particle density is $\bar{n} = \frac{1}{N}\sum_j\langle\hat{\sigma}_{A,j}^+\hat{\sigma}_{A,j}^-\rangle$.
    }
     \label{fig:CDW}
 \end{figure} 
%%%%%%%%%%%%%%%%%%%%%%%%%%%%%%%%%%%%%%%

To see why these single particle states emerge, consider the steady state, Eq.~\eqref{eq:ss-rewrite}, in the limit $|\zeta|\to0$: $|\psi_Q\rangle = e^{\sqrt{N}\hat{Q}/\drive^2}|\psi_\infty\rangle$.
In this limit, all holes in the state are created by some power of $\hat{Q}$ acting on $|\psi_\infty\rangle$, with the component $\hat{Q}^m|\psi_\infty\rangle$ having $2m$ holes.
For an $N$-length chain, $\hat{Q}$ can thus act up to $m = \lfloor N/2 \rfloor$ times to produce new states, after which $\hat{Q}^{m>\lfloor N/2 \rfloor}|\psi_\infty\rangle = 0$.
For an even chain, all $2m=N$ particles can be removed, taking the filled state to vacuum:
\begin{align}
    \hat{Q}^{\lfloor N/2 \rfloor}|\psi_\infty\rangle \propto |0\rangle,~~N~{\rm even}.
\end{align}
However, for an odd chain, all but 1 of the $N=2m+1$ particles can be removed.
The real-space pairing of holes on neighbouring sites requires that the remaining single $|\bullet\rangle$ particle is confined to the odd-$j$ sites, resulting in a CDW-like structure where a single particle is delocalized over the odd-$j$ sublattice only:
\begin{align}
    \hat{Q}^{\lfloor N/2 \rfloor}|\psi_\infty\rangle &\propto |\bullet\circ\circ\circ\cdots\rangle \\
    &\,- |\circ\circ\bullet\circ\cdots\rangle + \cdots,~~ N~{\rm odd}. \nonumber
\end{align}
Thus for $|\zeta| \ll 1$, there is a regime of sufficiently weak $|\drive|$ for which odd length chains exhibit a charge density wave consisting of a single delocalized particle.
The emergence of a CDW in an odd-length chain, and the lack of a CDW in an even-length chain for the same parameters, is shown in Fig.~\ref{fig:CDW}(a).
The particle density $\bar{n}=N|\drive|^4/8$ of the even chain is found using Eq.~\eqref{appeq:even-chain-no-cdw-density} in App.~\eqref{app:CDW-state-norms}.

The analysis follows analogously in the limit $|\zeta|\to\infty$, where now $|\psi_Q\rangle = e^{\sqrt{N}\hat{Q}/\drive^2}\hat{\tau}_1|\psi_\infty\rangle$ (as $\hat{\tau}_j$ commutes with $\hat{Q}$).
Site 1 thus always has a hole, and we repeat the above analysis on the remaining $N-1$ sites.
Thus, even chains have a CDW-like structure where a single particle is delocalized over the even-$j$ sites. 

The parameter regime in which CDWs emerge can be found by expanding $|\psi_Q\rangle$ to the lowest few orders in $|\drive|$.
Leaving the details to App.~\ref{app:CDW-state-norms}, we can show that two distinct scales emerge: a $N$-dependent upper limit on drive strength, $|\drive|^2\ll 1/N$, and a $|\zeta|$- and $N$-dependent lower limit that differs for even and odd length chains:
\begin{align}
    \frac{|\zeta|^2}{N} \ll |\drive|^2 \ll \frac{1}{N}, &\quad N~{\rm odd}, 
    \label{eq:cdw-regime-odd}\\
    \frac{1}{N|\zeta|^2} \ll |\drive|^2 \ll \frac{1}{N}, &\quad N~{\rm even}.
    \label{eq:cdw-regime-even}
\end{align}
When $|\drive|^2>1/N$, many particles can populate the chain, destroying the CDW ordering, and when $|\drive|^2<|\zeta|^{\pm2}/N$ (for $N$ odd/even), the average particle density vanishes as $\sim|\drive|^2$.
Here, Eq.~\eqref{eq:cdw-density-matrix} is no longer an equal mixture of $|0\rangle$ and $|\Phi_{\rm cdw}\rangle$, but increasingly weighted toward vacuum with $|\drive|\to0$.
The emergence of CDWs, and the dependence with both $N$ and $|\zeta|$, is shown in Fig.~\ref{fig:CDW}(b).

%%%%%%%%%%%%%%%%%%%%%%%%%%%%%%%%%%%%%%%%%%%%%%%%%%%%%%%%%%%%%%%%%%%%%%
%%%%%%%%%%%%%%%%%%%%%%%%%%%%%%%%%%%%%%%%%%%%%%%%%%%%%%%%%%%%%%%%%%%%%%

\section{Resource for cross-chain (remote) entanglement stabilization}
\label{sec:entanglement}

As discussed in Sec.~\ref{sec:setup}, the double qubit chain system in Fig.~\ref{fig:intro}(a) (where collective loss is provided by passive couplings to a waveguide) is a potentially powerful setup for stabilizing large amounts of steady-state remote entanglement.  
The scheme is also very resource efficient: it foregoes the complication and resource overhead of using squeezed light in favor of local driving, and it only requires passive hopping between qubit in each chain. It also does not require precise fine-tuning of parameters, nor does it require extremely strong driving to approach maximal entanglement of the chains.
We now use insights obtained from our exact solution Eq.~\eqref{eq:steady-state} to better understand this potential application.  

The exact solution tells us that for any drive strength $\Omega$, we will have a pure steady state with some degree of entanglement between the remote chain-$A$ and chain-$B$ qubits.  This entanglement is maximal in the $\Omega \rightarrow \infty$ limit, where the steady state becomes a dimerized product of cross-chain maximally entangled Bell pairs.  A natural question is {\it how strong} must the Rabi drive be to achieve this level of entanglement.  The exact solution provides a succinct and surprising answer here: one only needs that the {\it effective} drive amplitude $|\drive| \gtrsim 2$, as in this regime the density of ``holes" is very small, implying the steady state is very close to the ideal dimerized state.  We stress that this condition is independent of $N$ (even though drives are only applied to the first qubit in each change), and further, that one can achieve this condition even if $\Omega \ll \gamma$ (the drive does not need to overwhelm dissipation if hopping is sufficiently weak).   

Of course, these considerations neglect a crucial second issue: one cares both about the amount of entanglement in the dissipative steady state, as well as the time needed to prepare this state (i.e.~the characteristic system relaxation time, or the inverse dissipative gap).  This timescale will also directly determine the susceptibility of our scheme to additional unwanted dissipative processes (e.g.~waveguide loss, qubit dephasing and relaxation).  

A full study of the effects of waveguide loss and qubit dissipation on entanglement stabilization in a circuit QED realization of Fig.~\ref{fig:intro}(a) is presented in a complementary work \cite{irfan_Loss_2024}, but we briefly discuss the basic requirements to realize the scheme in circuit QED in App.~\ref{app:experiment}. Here, we instead focus on a fundamental aspect of the relaxation time physics in our double chain scheme.  While the qubit-only version suffers from a fundamental tradeoff between speed and entanglement, we show below that by generalizing the local 2-qubit scheme introduced in Ref.~\cite{brown_Trade_2022} 
to a multi-qubit setup with directional dissipation, we can dramatically improve this seemingly unavoidable tradeoff.  

%%%%%%%%%%%%%%%%%%%%%%%%%%%%%%%%%%%%%%%%%%%%%%%%%%%%%%%%%%%

\subsection{Slowdown in large drive limit}

Recent works \cite{motzoi_Backactiondriven_2016,govia_Stabilizing_2022,brown_Trade_2022,doucet_High_2020} on dissipatively preparing entanglement between two remote qubits have observed that the dissipative gap closes as the target entangled state approaches a perfect Bell pair (e.g., as the vacuum component of Eq.~\eqref{eq:steady-state-2qb} vanishes).
Refs.~\cite{brown_Trade_2022,doucet_High_2020} showed that this is a generic property of two-qubit systems, and that it arises due to an approximate conservation of total angular momentum that becomes an exact symmetry in the infinite driving limit.
In this limit, a second impure steady state emerges in the subspace orthogonal to the target Bell state.
As one approaches the point of added symmetry, the transition rate out of the orthogonal subspace and into the target Bell state becomes extremely small, leading to a vanishing dissipative gap.
We briefly review this argument in App.~\ref{app:slowdown}.

In the infinite driving limit $|\Omega/\Gamma|\to\infty$, the steady state of the $N=1$ double chain is not unique \cite{govia_Stabilizing_2022,doucet_High_2020,brown_Trade_2022}, but can be any state of the form $\hat{\rho}_1 = \nu|S\rangle \langle S| + \frac{1}{4}(1-\nu)\hat{\mathds{I}}$ for any $-1/3\leq \nu\leq 1$, see App.~\ref{app:steady-state-degeneracy} for details.
One can readily show that the maximally mixed state $\hat{\mathbb{I}}/4$ is replicated via the XX Hamiltonian.
Therefore, the near-symmetry that causes a slowdown in the $N=1$ system persists for $N>1$ because the near-steady infinite temperature state is replicated down the chain, and thus the chain cannot relax out of that state except by the very slow dissipative population transfer at the boundary.

\subsection{Speeding up stabilization with a qutrit}
\label{subsce:qutrit}

A recent work by Brown et al.~\cite{brown_Trade_2022} theoretically proposed and experimentally demonstrated that the slowdown associated with dissipatively stabilizing 2-qubit Bell pairs can be circumvented by promoting one of the qubits to a qutrit, in a system for which the dissipation is both local and reciprocal (i.e.
mediated by common coupling to a damped cavity mode).  They demonstrated that the near-symmetry that conserves total angular momentum is no longer present in a qubit-qutrit system. This makes the degenerate dark state vanish in the large drive limit. 

\begin{figure}[t!]
    \centering
    \includegraphics[width=0.98\columnwidth]{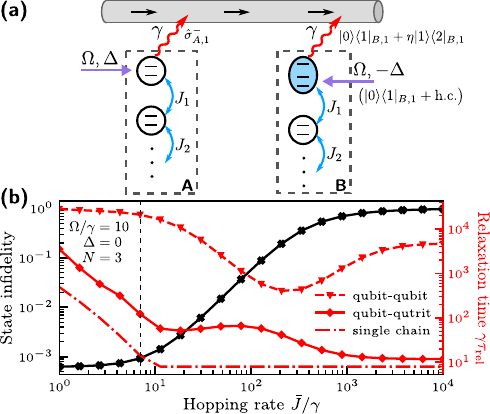}
    \caption{
        \textbf{Speeding up entanglement stabilization time of the nonreciprocal double chain using a qutrit.}
        \textbf{(a)}
        The nonreciprocal double chain is modified by replacing the downstream $B1$ qubit with a qutrit and engineering the nonreciprocal coupling to include the 2-1 transition of the qutrit.
        \textbf{(b)}
        The numerically computed relaxation time $\gamma\tau_{\rm rel}$ (red) and the infidelity of the steady state  to the maximally entangled state $|\psi_\infty\rangle$ (cf.~Eq.~\eqref{eq:steady-state-strong-drive}), $1- \langle\psi_\infty| \hat{\rho} |\psi_\infty\rangle$ (black), are shown as functions of the hopping rate $\Jbar/\gamma$ for $N=3$, $\Omega/\gamma = 10$, and $\Delta = 0$.
        For the qubit-qutrit scheme, the relaxation time is optimized over $\eta$.  We also plot the relaxation time for the single chain system for comparison.  
        For a fixed state fidelity of 0.999 (achieved at $\Jbar = 7\gamma$, dashed line), the relaxation times are $\gamma\tau_{\rm 2qb} = 2.1\times 10^4$, $\gamma\tau_{\rm qutrit} = 120$ and $\gamma\tau_{\rm sc} = 14$, for the qubit-qubit, qubit-qutrit, and single chain, respectively.
    }
    \label{fig:qutrit-scheme}
\end{figure}

Here, we show that this scheme can now be extended to the directional version of our double chain system by promoting the down-steam qubit $B1$ to a qutrit, see 
Fig.~\ref{fig:qutrit-scheme}(a). This leaves Eq.~\eqref{eq:steady-state} as the pure steady state while avoiding the symmetry-induced slowdown, and allows a dramatic stabilization speed up for arbitrary $N$ without sacrificing the fidelity with the perfect dimerized entangled state.  

More concretely, starting from the double chain master equation (cf.~Eq.~\eqref{eq:qme}) for $N=1$ in the directional limit $\nu=1$, we promote qubit $B1$ to a qutrit and modify its coupling to the waveguide via
\begin{align}
    \hat H_{\mathrm{drive}} &= \frac{\Omega}{2} \left(\sigma_{A,1}^x + |0 \rangle \langle 1 |_{B,1} + |1 \rangle \langle 0|_{B,1}  \right) \\
    &+ \frac{\Delta}{2}\left( \hat{\sigma}_{A,1}^z - \big[|1\rangle\langle 1|_{B,1} - |0\rangle\langle 0|_{B,1} \big]\right),\nonumber \\
    \hat H_{\mathrm{diss}} &= \frac{i \gamma}{2} \left( \sigma_{A,1}^+[|0 \rangle \langle 1 |_{B,1} + \eta |1 \rangle \langle 2 |_{B,1} ] - {\rm h.c.} \right), \\ 
    \hat L &= \sqrt{\gamma} \left( \sigma_{A,1}^- + |0 \rangle \langle 1 |_{B,1} + \eta |1 \rangle \langle 2 |_{B,1} \right), \label{eq:qutrit-master}
\end{align}
for which the master equation now reads $\partial_t \hat \rho = -i[\hat H_{\mathrm{drive}} + \hat H_{\mathrm{diss}}, \hat \rho] + \mathcal{D}[\hat L] \rho$.
Physically, this means that now the qutrit $B1$ can produce a photon in the chiral waveguide either via a $1-0$ relaxation event or a $2-1$ relaxation event (with relative matrix elements $\eta$).  
The result is a dissipative interaction that allows the state $|11 \rangle$ to pass a single photon through the waveguide at a rate $\eta \gamma$ and become $|02\rangle$, which can in turn decay into the state $|01 \rangle$. The effective interaction (no jump Hamiltonian) of such a process is $- i\eta \gamma \sigma_A^- (|2\rangle\langle 1|_B) $.  Because this process explicitly breaks the conservation of angular momentum in the 2 qubit subspace, it circumvents the slow down previously observed. 

From here, one can take our qubit-qutrit system and add back the remaining $N-1$ qubits in each chain, and the hopping Hamiltonian $\hat H_{\mathrm{XX}}$.  We stress that all remaining qubits are just qubits:  it is only $B1$ where we need to make use of the higher $|2\rangle$ level.  A direct calculation shows that the steady state found in Eq.~\eqref{eq:steady-state} is still a zero-energy eigenstate of the new $\hat{H}_{\rm drive} + \hat H_{\mathrm{diss}}$, as well as the new jump operator $\hat L$, and so it remains a dissipative steady state; the only thing that has changed in adding the third level is the dynamics, which should now be significantly faster. This is demonstrated in Fig.~\ref{fig:qutrit-scheme}(b) for an $N=3$ system.

Numerically, we observe a significant improvement in the relaxation time scale $\tau_{\rm rel}$ (as determined by the inverse dissipative gap of the full Lindbladian) of a $N=3$ system, when we promote site $B1$ to a qutrit and optimize over the qutrit 2-1 transition coefficient $\eta$.  These results are as shown in Fig.~\ref{fig:qutrit-scheme}(b).
Here, we fix all other parameters except the uniform hopping rate $\Jbar$, and show how both the fidelity with the ideal dimerized entangled state and $\tau_{\mathrm{rel}}$ vary with $\Jbar$.  
For the all-qubit chain, small $\Jbar$ makes $|\drive|$ large, thus the fidelity to the maximally entangled state $|\psi_\infty\rangle$ (cf.~Eq.~\eqref{eq:steady-state-strong-drive}) is high, but the relaxation slows down.
As Fig.~\ref{fig:qutrit-scheme}(b) shows, there is a dramatic improvement of the relaxation time: over two orders of magnitude at a state fidelity of 0.999.
The optimized value of $\eta$ as a function of $\Jbar/\gamma$, as well as the speed-up of an $N=2$ system, is shown in App.~\ref{app:optimization-eta}.
We expect that the qutrit scheme speeds up the stabilization time for larger $N$ systems as well.

%%%%%%%%%%%%%%%%%%%%%%%%%%%%%%%%%%%%%%%%%%%%%%%%%%%%%%%%%%%%%%%%%%%%%%
%%%%%%%%%%%%%%%%%%%%%%%%%%%%%%%%%%%%%%%%%%%%%%%%%%%%%%%%%%%%%%%%%%%%%%

\section{Conclusion}
\label{sec:conclusion}

Our work presents an exact analytic solution for the steady state of two different spin chain models with boundary dissipation and driving.  As discussed, these solutions reveal a number of surprising correlation effects (e.g.~the effective pairing of holes), and lay the groundwork for a potentially powerful route to dissipative stabilization of remote multi-qubit entanglement.  We also elucidated a general mechanism for ``replicating'' definite-parity two-qubit entangled states using passive XX couplings in a double qubit chain, and demonstrated that the approach of Ref.~\cite{brown_Trade_2022} for avoiding slowdowns in dissipative entanglement stabilization could be extended from a 2 qubit situation to a setup with many qubits and directional dissipation.     

In future work, it will be extremely interesting to explore whether the ideas introduced here could be extended to more complex systems, where multiple $1$D XX qubit chains are attached to the same common waveguide.  This could potentially be a source of stabilized multi-partite, multi-qubit remote entanglement.  It would also be interesting to explore further the dynamics of our solvable dissipative spin chain models.  As discussed, the solvability of the non-trivial single chain model in Fig.~\ref{fig:intro}(a) can be ultimately traced to a surprising hidden time-reversal symmetry \cite{roberts_Hidden_2021}.  Understanding how this symmetry constrains the dynamics and Liouvillian spectrum could be an extremely rich direction for future research.  It would also be interesting to understand whether the scaling of the dissipative gap in the two chain model of Fig.~\ref{fig:intro}(b) could be improved beyond the usual $1/N^3$ scaling that is found in a variety of integrable spin chain models \cite{znidaric_Relaxation_2015, pocklington_Stabilizing_2022}.  
Finally, it would be interesting to study the spectra and NESS of other two-chain models. We already demonstrated in App.~\ref{app:rsga-scar} that our hole pairing states correspond to many-body scar states in a closed, non-intergrable ladder system.  Adding dissipation here could be extremely interesting.  Further,  one could extend our two-chain model to a ladder system with Creutz ladder style couplings \cite{alaeian_Creating_2019} along the full length of the chain.  
If one tunes the diagonal interchain couplings $t_d$ to be equal to the intra-chain XX couplings $J$, then this system possesses at least $N-1$ strong symmetries.  It thus has multiple dissipative steady states, making it another interesting system worthy of further study.

\emph{Acknowledgements.} We thank P. Rabl and J. Agustí for useful discussions.
We also thank the anonymous referee who pointed out the connection of our hole pairing construction to quantum scars,  and the existence of strong symmetries in a modified model with Creutz ladder style couplings.
This work was supported by the National Science Foundation QLCI HQAN (NSF Award No.~2016136).  AL, AP, MY, YXW and AC acknowledge support from the Army Research Office under Grant No.~W911NF-23-1-0077, and from the Simons Foundation through a Simons Investigator Award (Grant No.~669487).  
This work was completed in part with resources provided by the University of Chicago’s Research Computing Center.

\emph{Note added:} While completing our manuscript, we became aware of a related but independent work on autonomously stabilizing many-qubit entanglement; unlike our study, the setup in this work used squeezed light and explicitly directional qubit-qubit couplings \cite{agusti_Autonomous_2023}.

%%%%%%%%%%%%%%%%%%%%%%%%%%%%%%%%%%%%%%%%%%%%%%%%%%%%%%%%%%%%%%%%%%%%%%
%%%%%%%%%%%%%%%%%%%%%%%%%%%%%%%%%%%%%%%%%%%%%%%%%%%%%%%%%%%%%%%%%%%%%%


\newpage

\appendix

\section{Interacting fermion model of the dissipative spin chain}
\label{app:single-spin-chain}

Given a boundary driven/dissipative spin chain, often the first step towards obtaining a solution is performing a Jordan-Wigner transform into free fermions \cite{jordan_Ueber_1928}. Define the canonical (Dirac) fermions:
\begin{align}
\hat c_j &= \left( \prod_{i = 1}^{j-1} \hat \sigma_i^z \right) \hat \sigma_j^-.
\end{align}
Then we can rewrite the spin Hamiltonian [c.f.~$\hat H_A$ in Eq.~\eqref{eq:single-chain-qme}] as
\begin{align}
\hat H_{\mathrm{fermi}} &= \frac{\Omega}{2} (\hat c_1 + \hat c_1^\dagger) + \Delta \hat c_1^\dagger \hat c_1 - \frac{1}{2} \sum_{j=1} ^{N-1} J_j \left( \hat c_j^\dagger \hat c_{j + 1} + {\rm h.c.} \right). \label{eqn:fermi_ham}
\end{align}
The Hamiltonian is a sum of quadratic and linear fermion terms, and can be exactly diagonalized (using the procedure outlined in \cite{colpa_Diagonalisation_1979}). 

We are of course interested in the  dissipative dynamics.  The full master equation in terms of fermions is then
\begin{align}
\dot{\hat{\rho}} &= -i[\hat H_{\mathrm{fermi}}, \hat \rho] + \gamma \mathcal{D} [\hat c_1] \hat \rho. \label{eqn:fermi_master}
\end{align}
It also has both quadratic and linear fermion terms. While one might assume that this master equation is exactly solvable, this is not the case.  
If the fermionic Lindbladian contains \textit{both} a Hamiltonian term linear in fermion operators, along with linear dissipation, then this will generically correspond to an interacting problem. The easiest way to see this is simply to compute the equation of motion for the linear expectation value of a fermionic operator under just the dissipative dynamics. Because the Hamiltonian has a linear term, the even and odd moments are dynamically connected and so we must consider these linear expectation values.
\begin{align}
\partial_t \langle \hat c_j \rangle &= \gamma \left\langle \mathcal{D}[\hat c_1]^\dagger \hat c_j \right\rangle= \frac{\gamma}{2} \left\langle \hat c_1^\dagger [\hat c_j, \hat c_1] + [ \hat c_1^\dagger, \hat c_j] \hat c_1 \right\rangle, \\
&= -\frac{\gamma}{2} \left\langle 4 \hat c_1^\dagger \hat c_1 \hat c_j + \delta_{1,j} \hat c_j \right\rangle.
\end{align}
We see that these first moments are coupled to third moments.  
In a similar manner, one can observe that all odd moments of degree $n$ couple to odd moments of degree $n + 2$, and so the equations of motions do not close on themselves, signifying that this is not a simple free fermion model. 

 \begin{figure}[t]
    \centering
    \includegraphics[width=0.98\columnwidth]{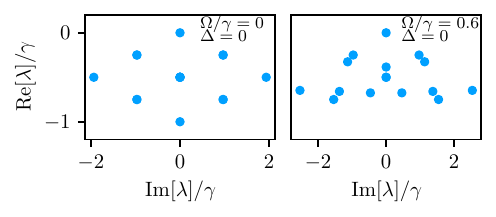}
     \caption{
        \textbf{Lindblad spectrum of a single chain with 2 qubits.}
        We plot the eigenvalues $\lambda$ of the single chain master equation Eq.~\eqref{eq:single-chain-qme} for $N=2, \Delta = 0$ and $J = \gamma$.  Left:  In the absence of any drive $\Omega = 0$, the spectrum has the normal mode form expected for a quadratic fermionic Lindblad master equation.  Right:  For a non-zero drive $\Omega = 0.6 \gamma$, the normal mode structure is lost, e.g.~summing single-excitation eigenvalues does not correctly predict higher lying eigenvalues.  This provides a direct confirmation that with driving and loss, the single chain problem is not equivalent to a quadratic fermionic Lindbladian.
    }
     \label{fig:spectrum}
 \end{figure}

More formally, one could try and use the standard diagonalization technique for a linear fermionic Hamiltonian \cite{colpa_Diagonalisation_1979} and introduce a fictitious fermion $\hat \eta$ to homogenize the Hamiltonian and make everything quadratic. This is equivalent to rewriting the Hamiltonian Eq.~\eqref{eqn:fermi_ham} as 
\begin{align}
\hat H_{\mathrm{fermi, \eta}} &= \frac{\Omega}{2} (\hat c_1 + \hat c_1^\dagger)(\hat \eta - \hat \eta^\dagger) + \Delta \hat c^\dagger \hat c \nonumber \\
& \ \ \ \ - \frac{1}{2} \sum_{j=1} ^{N-1} J_j \left( \hat c_j^\dagger \hat c_{j + 1} + {\rm h.c.} \right).
\end{align}
Now, the Majorana $(\hat \eta + \hat \eta^\dagger)$ is conserved by the Hamiltonian, and so if this were a closed system we would be done as the two Hamiltonians $\hat H_{\mathrm{fermi, \eta}}$ and $\hat H_{\mathrm{fermi}}$ would be isospectral (with $\hat H_{\mathrm{fermi, \eta}}$ doubly degenerate). However, this is an open system, and we need to consider the dissipation. In this case, the Majorana
$(\hat \eta + \hat \eta^\dagger)$
only constitutes a \textit{weak} symmetry \cite{buca_note_2012} of the Lindbladian as it anticommutes with the jump term:
\begin{align}
\{ \hat \eta + \hat \eta^\dagger, \hat c_1 \} &= 0.
\end{align}
This is problematic, as without further modification, the dissipation will cause unphysical jumps between the two conserved sectors of $\hat{H}_{\mathrm{fermi},\eta}$.

To correct this problem, we must also modify the linear-in-fermion jump operator so that these unphysical jumps do not occur.  Formally, we must make the conservation of $(\hat \eta + \hat \eta^\dagger)$ a strong symmetry \cite{buca_note_2012}.  This is achieved by the master equation
\begin{align}
\dot{\hat{\rho}} &= -i[\hat H_{\mathrm{fermi,\eta}}, \hat \rho] + \gamma \mathcal{D} [(-1)^{\hat \eta^\dagger \hat \eta}\hat c_1] \hat \rho.
\end{align}
Note that this is ultimately equivalent to first introducing an auxilliary \textit{spin} in Eq.~\eqref{eq:single-chain-qme} (preceding the first lattice site), and then performing the Jordan-Wigner transform. Note that the jump operator is now cubic, and therefore the system is explicitly interacting. We also note that this procedure is consistent with the general rules outlined in  Ref.~\cite{colpa_Diagonalisation_1979}:  when introducing the auxiliary fermion $\eta$, {\it all} linear-in-fermion operator terms  must be modified to ensure that they have the correct matrix elements in the expanded space.  This rule must be applied to the jump operator $\hat{c}_1$, as the action of the superoperator  $\mathcal{D}[\hat{c}_1]$ cannot be written solely in terms of the quadratic operator $\hat{c}^\dagger_1 \hat{c}_1$.    

As a final confirmation that the single-chain qubit system is not equivalent to free fermions, in Fig.~\ref{fig:spectrum}, we plot the full Liouvillian spectrum (eigenvalues $\lambda$) for the $N=2$ version of the master equation Eq.~\eqref{eq:single-chain-qme}  For $\Omega=0$ (left panel), the eigenvalues have the normal mode structure expected for a free fermion Lindblad master equation (as can be found using third quantization \cite{prosen_Third_2008}).  This implies, e.g., that eigenvalues corresponding to two particle excitations are formed by summing eigenvalues associated with single particle excitations.  With a non-zero drive (right panel), this structure is clearly lost.  The Lindblad spectrum with both drive and loss no longer has the form expected for a free fermion Lindbladian, consistent with our conclusions above.

%%%%%%%%%%%%%%%%%%%%%%%%%%%%%%%%%%%%%%%%%%%%%%%%%%%%%%%%%%%%%%%%%%%%%%
%%%%%%%%%%%%%%%%%%%%%%%%%%%%%%%%%%%%%%%%%%%%%%%%%%%%%%%%%%%%%%%%%%%%%%

\section{Replication}
\subsection{Intuition for fixed parity states}
\label{app:XXintuition}

The fact that the passive XX couplings can replicate fixed parity states is simple enough to check, but is not immediately intuitive. To get some more intuition, let's write the Hamiltonian for a pair of dimers as
\begin{align}
\hat H &= \hat\sigma^x_{A,1} \hat\sigma^x_{A,2} + \hat\sigma^y_{A,1} \hat\sigma^y_{A,2} - \left( \hat\sigma^x_{B,1} \hat\sigma^x_{B,2} + \hat\sigma^y_{B,1} \hat\sigma^y_{B,2} \right)
\end{align}
We can immediately observe that $[\hat H, \hat S^2_{A,B}] = [\hat H, \hat S^z_{A,B}] = 0$, and so we can diagonalize it in terms of the singlet and triplet states $|s_A,m_A \rangle \otimes |s_B, m_B \rangle$ where $s,m$ are the standard quantum numbers of total spin and z-angular momentum. It is quite easy to observe that
\begin{widetext}
\begin{align}
( \hat\sigma^x_1 \hat\sigma^x_{2} + \hat\sigma^y_{1} \hat\sigma^y_{2})|s,m\rangle &= \delta_{m,0} (-1)^{s + 1} |s,m\rangle \\
\implies \hat H |s_A,m_A \rangle \otimes |s_B, m_B \rangle &= \left( \delta_{m_A,0} (-1)^{s_A + 1} - \delta_{m_B,0} (-1)^{s_B + 1} \right)|s_A,m_A \rangle \otimes |s_B, m_B \rangle \\
\implies \hat H |s_A,m_A \rangle \otimes |s_B, m_B \rangle &= 0
\end{align}
\end{widetext}
The only remaining piece of the puzzle is the observation that tensoring together two copies of a fixed (even) parity state $|\psi \rangle$ is diagonal in this spin basis.
\begin{align}
|\psi \rangle &= \sum_{i \in \{0,1\}} \psi_i |i \rangle_A \otimes |i \rangle_B \\
\implies |\psi \rangle_1 \otimes |\psi \rangle_2 &= \sum_{i,j \in \{0,1\}} \psi_i \psi_j |i j \rangle_A \otimes |i j \rangle_B \label{eqnA:comp} \\
&= \sum_{s,m} c_{s,m} |s,m \rangle \otimes |s,m \rangle \label{eqnA:spin},
\end{align}
where in the last line we go from the computational basis in Eq.~\eqref{eqnA:comp} to the total spin basis in Eq.~\eqref{eqnA:spin}. Since the coefficient $\psi_i \psi_j$ is a symmetric tensor, it is diagonal in the spin basis; intuitively, this is because the spin states are all either symmetric ($s = 1$) or antisymmetric ($s=0$). Thus, the tensor product of $s = 0$ with $s = 1$ is antisymmetric, which multiplied by a symmetric tensor is identically zero. However, the product of two symmetric or two antisymmetric tensors won't be. This can also be checked very straightforwardly by direct computation. $|00\rangle$ in the computational basis is $|1,-1\rangle$ in the total spin basis already, and vice versa for $|11\rangle \leftrightarrow |1,1 \rangle$. Thus, it only requires checking the cross terms. Denote $|S\rangle = |0,0\rangle = (|01\rangle - |10\rangle)/\sqrt{2}$ the singlet state, and $|T\rangle = |1,0\rangle = (|01\rangle + |10\rangle)/\sqrt{2}$ the triplet. Then
\begin{align}
    &|01\rangle \otimes |01\rangle + |10 \rangle \otimes |10 \rangle \nonumber \\
    &= \frac{1}{2}(|S\rangle + |T\rangle) \otimes (|S\rangle + |T\rangle) + \frac{1}{2} (|S\rangle - |T\rangle) \otimes (|S\rangle - |T\rangle) \nonumber \\
    &= |S\rangle \otimes |S\rangle + |T \rangle \otimes |T\rangle
\end{align}
as expected.

Going to a fixed (odd) parity state can be deduced in the same manner by observing that (defining a bit flip operation via $|\overline{i} \rangle \equiv |1 - i\rangle 
= \hat \sigma ^{x}
|i\rangle $)
\begin{align}
|\psi \rangle &= \sum_i \psi_i |i \rangle_A \otimes |\overline{i} \rangle_B \\
\implies |\psi \rangle_1 \otimes |\psi \rangle_2 &= \sum_{i,j} \psi_i \psi_j |i j \rangle_A \otimes |\overline{i} \overline{j} \rangle_B \\
&= \sum_{s,m} c_{s,m} |s,m \rangle \otimes |\overline{s,m }\rangle.
\end{align}
However, bit flip on the entire $B$ chain leaves the Hamiltonian invariant:
\begin{align}
\hat\sigma^x_{B,1} \hat\sigma^x_{B,2}
 \hat H
\hat\sigma^x_{B,1} \hat\sigma^x_{B,2}= \hat H
\end{align}
and hence the argument still holds.

This shows that the XX Hamiltonian annihilates every tensor product of identical fixed parity states. Thus, we can repeat this argument for an arbitrarily long chain of identical, fixed parity states. If we denote $\hat H_{{\rm XX}, i}$ as the XX Hamiltonian acting on the $i$ and $i + 1$ pair of spins, then the total Hamiltonian would be 
\begin{align}
    &\hat H_{\mathrm{total}} |\Psi \rangle_{\mathrm{total}} = \left( \sum_{i = 1}^{N - 1} J_i \hat H_{{\rm XX}, i} \right)|\psi \rangle^{\otimes N} \nonumber \\
    =& \sum_{i = 1}^{N-1} |\psi\rangle_1 \dots \left( J_i \hat H_{{\rm XX},i} |\psi\rangle_i |\psi \rangle_{i + 1} \right) \dots |\psi \rangle_N \\ 
    =& 0,
\end{align}
so the total $2N$-site XX Hamiltonian has as a many-body zero-energy eigenstate composed of $N$ copies of an arbitrary fixed parity 2-qubit state.

\subsection{Heisenberg Couplings}
\label{app:heisreplication}

To observe that Heisenberg couplings can replicate any state, it is first important to observe that, given a fixed (even) parity state $|\psi \rangle = u|00\rangle + v |11 \rangle$, then
\begin{align}
    \left(\hat\sigma_{A,1}^z \hat\sigma_{A,2}^z - \hat\sigma_{B,1}^z \hat\sigma_{B,2}^z \right) |\psi \rangle \otimes |\psi \rangle  = 0
\end{align}
Combining this with the fact that, as shown in the previous section, this state is annihilated by the XX couplings, we see that it is annihilated by any $XXZ$ Hamiltonian.

From here, we point out that given an arbitrary state $|\phi \rangle$, then by the Schmidt decomposition, there exist local unitaries $\hat U_1, \hat U_2$ such that
\begin{align}
   \left(  \hat U_1 \otimes \hat U_2 \right) |\phi \rangle &= u |00 \rangle + v |11 \rangle \equiv |\psi \rangle
\end{align}
for some $u,v$. Let's define $\hat H_H$ to be the isotropic Heisenberg Hamiltonian. Then we have that
\begin{align}
    \hat H_H |\phi \rangle \otimes |\phi \rangle &= \hat H_H \left( U_{1}^\dagger \otimes U_{1}^\dagger \right)_A \otimes \left( U_{2}^\dagger \otimes U_{2}^\dagger \right)_B |\psi \rangle \otimes |\psi \rangle 
\end{align}
Since the isotropic Heisenberg Hamiltonian is invariant under uniform local unitary rotations, it can be commuted through to annihilate the fixed parity state:
\begin{align}
     \hat H_H |\phi \rangle \otimes |\phi \rangle &= 0
\end{align}
as desired.

\subsection{Replication in more complicated graphs}
\label{app:replication-trees}

We will now demonstrate the claim made in the main text that the replication mechanism can work on more complicated graphs than a 1D chain. In fact, states can be replicated down any tree-like structure (i.e., with no closed loops) that has exactly one symmetry axis.

The proof is a simple generalization of what we have already shown: we will show that, given two dissipatively stabilized qubits, one can attach \textit{an arbitrary} number of qubit pairs off of these using symmetric XX couplings. This generates one level of the tree graph, and then simple bootstrapping shows that arbitrary trees are possible.

Let's assume once again that there exists a Liouvillian operator $\mathcal{L}_0$ acting on qubits at site $A_0, B_0$ that stabilizes a fixed parity state $|\psi\rangle$, $\mathcal{L}_0(|\psi \rangle \langle \psi |) = 0$. Next, we will extend this to the next layer in the graph by defining the next set of qubits on sites $A_1, \dots, A_n$ and $B_1, \dots, B_n$, so that the full system Liouvillian is now
\begin{align}
    \mathcal{L} &= \mathcal{L}_0 \otimes \mathds{1}^{\otimes n} -i[\hat H_{{\rm XX},n}, \cdot] \\
    \hat H_{{\rm XX},n} &= \sum_{s = A,B} \sum_{i = 1}^n \mathrm{sgn}(s) J_i \left( \sigma_{s,0}^x \sigma_{s,i}^x +\sigma_{s,0}^y \sigma_{s,i}^y \right)
\end{align}
where we define $\mathrm{sgn}(A) = 1, \mathrm{sgn}(B) = -1$. Now at this point, defining $|\Psi \rangle = |\psi \rangle^{\otimes n + 1}$, it is simple to observe that since the Hamiltonian is simply a sum of terms acting independently on the different dimers, $\hat H_{{\rm XX},n} |\Psi \rangle = 0$ by the exact same logic as presented before.

Thus, a single qubit can sustain any number of pairs branching off of it. However, this means that each of those are now dissipatively stabilized fixed parity states, and so we can repeat the argument to branch more qubits off, generating trees.

At this point, it is crucial to note that, since there are multiple qubits branching off a single pair, now it is important that the $J_i$ terms are all distinct if you want a unique steady state. If there are degeneracies in the $J_i$ parameters, then one generates a permutation symmetry where there are multiple degenerate pairings between $A$ qubits and $B$ qubits, and so the steady state will necessarily be degenerate. 

%%%%%%%%%%%%%%%%%%%%%%%%%%%%%%%%%%%%%%%%%%%%%%%%%%%%%%%%%%%%%%%%%%%%%%
%%%%%%%%%%%%%%%%%%%%%%%%%%%%%%%%%%%%%%%%%%%%%%%%%%%%%%%%%%%%%%%%%%%%%%

\section{Dimer representation}
\label{app:dimer-rep}

Each dimer chain site can be described as a pair of non-commuting spin-1's embedded in the spin-$\frac{1}{2}$$\times$spin-$\frac{1}{2}$ Hilbert space which share $m=\pm1$ states but have orthogonal $m=0$ states.
The $m=-1$ state is the two-qubit vacuum $|00\rangle$, the $m=+1$ state is the double-excited state $|11\rangle$, and the orthogonal pair of $m=0$ states are $|S\rangle=(|01\rangle-|10\rangle)/\sqrt{2}$ and $|T\rangle=(|01\rangle+|10\rangle)/\sqrt{2}$. 

We define the lowering operators $\hat{\tau}_j$ and $\hat{\lambda}_j$ on each site $j$ that each destroy one of $|S_j\rangle$ or $|T_j\rangle$. The $\hat{\tau}_j$ destroy  $|S_{j=1,3,5,\dots}\rangle$ and $|T_{j=2,4,6,\dots}\rangle$ -- the $|\bullet\rangle$ particles -- and the $\hat{\lambda}_j$ destroy the opposite $m=0$ states -- the $|{\sqr}\rangle$ particles.
Explicitly, in terms of the qubit operators, the lowering operators are:
\begin{align}
    \hat{\tau}_{j} &\equiv \hat{\sigma}_{B,j}^{-}+(-1)^{j}\hat{\sigma}_{A,j}^{-} \label{appeq:tau-op}\\
    &= \sqrt{2} 
    \begin{cases}
        |00_{j}\rangle\langle S_{j}| - |S_{j}\rangle\langle 11_{j}| & j~{\rm odd} \\
        |00_{j}\rangle\langle T_{j}| - |T_{j}\rangle\langle 11_{j}| & j~{\rm even}
    \end{cases}, \nonumber
\end{align}
\begin{align}
    \hat{\lambda}_{j} &\equiv \hat{\sigma}_{B,j}^{-}-(-1)^{j}\hat{\sigma}_{A,j}^{-} \label{appeq:lambda-op}\\
    &= \sqrt{2} 
    \begin{cases}
        |00_{j}\rangle\langle S_{j}| + |S_{j}\rangle\langle 11_{j}| & j~{\rm even} \\
        |00_{j}\rangle\langle T_{j}| + |T_{j}\rangle\langle 11_{j}| & j~{\rm odd}
    \end{cases}. \nonumber
\end{align}
Each ladder operator and its adjoint forms a spin-1 representation of SU(2), with the same $\hat{S}^z_j$ completing the algebra for each. That is, the commutation relations $[\hat{\tau}_{j}^{\dagger},\hat{\tau}_{k}]=[\hat{\lambda}_{j}^{\dagger},\hat{\lambda}_{k}]=2\hat{S}^{z}_{j}\delta_{jk}$ are simultaneously satisfied for the operator
\begin{align}
    \hat{S}^{z}_{j} = \frac{1}{2}\left(\hat{\sigma}_{A,j}^{z} + \hat{\sigma}_{B,j}^{z} \right) = |11_{j}\rangle\langle 11_{j}| - |00_{j}\rangle\langle 00_{j}|.
\end{align}
The two flavors of ladder operators on each site $j$ do not commute with each other:
\begin{align}
    [\hat{\tau}^{\dagger}_{j},\hat{\lambda}_{j}] = [\hat{\lambda}^{\dagger}_{j},\hat{\tau}_{j}] &= \hat{\sigma}_{B,j}^{z} - \hat{\sigma}_{A,j}^{z}
    \label{eq:flavor-commutator} \\
    &= 2\left( |S_j\rangle\langle T_j| + {\rm h.c.} \right). \nonumber
\end{align}
Notice that these commutators are Hermitian and act to change flavor on a given dimer site. Finally products of same-site ladder operators are:
\begin{align}
    \hat{\tau}_{j}^{2} &= -\hat{\lambda}_{j}^{2} = -2(-1)^{j} |00_{j}\rangle\langle 11_{j}|,\\
    \hat{\tau}_{j}\hat{\lambda}_{j} &=0, \label{appeq:flavor-orthogonality}
\end{align}
This completes the dimer algebra.

We use the following notation of states in the dimer representation: $|\circ_{j}\rangle$ and $|(11)_{j}\rangle$ are the vacuum and double-excited state, respectively. The single-excited states are $|\bullet_{j}\rangle = \frac{1}{\sqrt{2}}\hat{\tau}^{\dagger}_{j}|\circ_{j}\rangle$ and $|{\sqr}_{j}\rangle = \frac{1}{\sqrt{2}}\hat{\lambda}^{\dagger}_{j}|\circ_{j}\rangle$.
Finally, we define the steady-state subspace 
\begin{align}
    \mathcal{H}_s = {\rm span}\,\Big\{\sum_j|\circ_j\rangle,|\bullet_j\rangle\Big\}
    \label{appeq:Hs}
\end{align}
(so-called because $|\psi_Q\rangle\in\mathcal{H}_s$) which contains all states with only $|{\bullet}_j\rangle$ and $|{\circ}_j\rangle$.
The projector into $\mathcal{H}_s$ is given by
\begin{align}
    \proj = \bigotimes_{j} \left( |\circ_{j}\rangle\langle \circ_{j}| + |\bullet_{j}\rangle\langle \bullet_{j}| \right).
    \label{appeq:proj}
\end{align}

Now we rewrite Eq.~\eqref{eq:qme} in the dimer representation.
The collective loss operator (cf. Eq.~\eqref{eq:closs}) is simply
\begin{align}
    \hat{c} = \hat{\lambda}_{1},
    \label{eq:clossdimer}
\end{align}
which makes immediately obvious that its dark subspace on site 1 is spanned by $|\circ_{1}\rangle$ and $|\bullet_{1}\rangle$ since $\hat{\lambda}_1|(11)_1\rangle = \sqrt{2}|{\sqr}_1\rangle$ and $\hat{\lambda}_1|{\sqr}_1\rangle = \sqrt{2}|\circ_1\rangle$.
The Hamiltonian terms (cf. Eqs.~\eqref{eq:Hdrive},\eqref{eq:Hxx}, and \eqref{eq:Hdiss}) are
\begin{align}
    \hat{H}_{\rm XX} &= \frac{1}{4}\sum_{j=1}^{N-1} J_j\left( \hat{\tau}^{\dagger}_{j}\hat{\lambda}_{j+1} + \hat{\lambda}^{\dagger}_{j}\hat{\tau}_{j+1} + {\rm h.c.} \right), 
    \label{appeq:Hxx-dimer}\\
    \hat{H}_{\rm drive} &= \frac{\Omega}{2} \left(\hat{\lambda}^{\dagger}_{1} +{\rm h.c.} \right) - \frac{\Delta}{2} \big[ \hat{\lambda}^{\dagger}_{1},\hat{\tau}_{1} \big], 
    \label{appeq:Hdrive-dimer}\\
    \hat{H}_{\rm diss} &= \frac{1}{4}i\nu\gamma \left(\hat{\lambda}^{\dagger}_{1}\hat{\tau}_{1} - {\rm h.c.} \right).
    \label{appeq:Hdiss-dimer}
\end{align}
$\hat{H}_{\rm XX}$ becomes a flavor-changing exchange interaction and the Rabi drives become a single drive acting on the $|{\sqr}\rangle$ flavor.
The drive detuning in $\hat{H}_{\rm drive}$ and the nonreciprocity-induced exchange in $\hat{H}_{\rm diss}$ appear in different ways to achieve the same effect: a flavor change that swaps $|\bullet_1\rangle$ and $|{\sqr}_1\rangle$.
Note that the commutator in $\hat{H}_{\rm drive}$ is Hermitian, which can be seen using Eq.~\eqref{eq:flavor-commutator}.

%%%%%%%%%%%%%%%%%%%%%%%%%%%%%%%%%%%%%%%%%%%%%%%%%%%%%%%%%%%%%%%%%%%%%%
%%%%%%%%%%%%%%%%%%%%%%%%%%%%%%%%%%%%%%%%%%%%%%%%%%%%%%%%%%%%%%%%%%%%%%

\section{Hole-pairing and {\it XX} eigenstates in qubit chains and Fermi-Hubbard chains}
\label{app:hole-pairing}

\subsection{Hole-pairing operator}

The hole-pairing operator $\hat{Q}$ defined by Eq.~\eqref{eq:Qpair} is a central character in the analytical description of the steady state of Eq.~\eqref{eq:qme}.
Here, we prove that it acts on eigenstates of $\hat{H}_{\rm XX}$ (cf. Eq.~\eqref{eq:Hxx}) within the steady-state subspace $\mathcal{H}_s$ (cf. Eq.~\eqref{appeq:Hs}) to produce new eigenstates of $\hat{H}_{\rm XX}$ with the same energy and also in $\mathcal{H}_s$.
We prove this at the operator level by showing that when $\hat{Q}$ acts on states within the steady-state subspace $\mathcal{H}_s$, it commutes with both $\hat{H}_{\rm XX}$ and the projector into the subspace $\proj$ (cf. Eq.~\eqref{appeq:proj}),
\begin{align}
    [\hat{Q},\hat{H}_{\rm XX}]\proj &= 0, \\
    [\hat{Q},\proj]\proj &= 0.
\end{align}
Together, these imply that for a given $\hat{H}_{\rm XX}$ eigenstate $|\phi\rangle\in\mathcal{H}_{s}$ with energy $E$, the action of $\hat{Q}$ on $|\phi\rangle$ produces another eigenstate with energy $E$:
\begin{align}
    (\hat{H}_{\rm XX}-E)(\hat{Q}|\phi\rangle) &= 0,\quad \hat{Q}|\phi\rangle \in \mathcal{H}_s.
\end{align}
In this way, we find that by repeated application of $\hat{Q}$ on an $\hat{H}_{\rm XX}$ eigenstate, a tower of some finite number $n_0<\infty$ of degenerate eigenstates is returned.

In general the commutators $[\hat{Q},\hat{H}_{\rm XX}] \neq 0$ and $[\hat{Q},\proj] \neq 0$.
A direct computation using Eqs.~\eqref{appeq:proj} and \eqref{appeq:Hxx-dimer} yields the commutators
\begin{widetext}
    \begin{align}
        [\hat{Q},\hat{H}_{\rm XX}] &=\frac{1}{8\sqrt{N}}\sum_{j}(-1)^{j}\left[\frac{J_{j}^{2}}{\bar{J}}[\hat{\tau}_{j},\hat{\lambda}_{j}^{\dagger}]\left(\hat{\tau}_{j+1}^{2}-\hat{\tau}_{j-1}^{2}\right)-2\frac{J_{j}J_{j-1}}{\bar{J}}\hat{S}_{j}^{z}\left(\hat{\tau}_{j+1}\hat{\lambda}_{j-1}-\hat{\tau}_{j-1}\hat{\lambda}_{j+1}\right)\right],
        \label{appeq:Q-Hxx-comm}\\
        [\hat{Q},\proj]&=-\frac{1}{\sqrt{N}}\sum_{j}(J_{j}/\bar{J})(-1)^{j}\bigotimes_{k\neq j,j+1}\left(|\circ_{k}\rangle\langle\circ_{k}|+|\bullet_{k}\rangle\langle\bullet_{k}|\right) \times\\
        &\qquad\qquad\qquad\Big[|{\bullet}_{j}\circ_{j+1}\rangle\langle(11)_{j}{\bullet}_{j+1}|+|{\circ}_{j}{\bullet}_{j+1}\rangle\langle{\bullet}_{j}(11)_{j+1}|+|{\bullet}_{j}{\bullet}_{j+1}\rangle\langle(11)_{j}(11)_{j+1}|\Big]. \nonumber
    \end{align}
\end{widetext}
Noting that $\hat{\tau}_j^2|\bullet_j\rangle = \hat{\lambda}_j|\bullet_j\rangle=0$ and $\hat{\tau}_j^2|\circ_j\rangle = \hat{\lambda}_j|\circ_j\rangle=0$, we immediately see that within the steady-state subspace, $[\hat{Q},\hat{H}_{\rm XX}]\proj = 0$.
By inspection, $[\hat{Q},\proj]\proj=0$.

\subsection{Hole-pairing in a Fermi-Hubbard model}

The hole-pairing observed in the dimer chains is not unique to that system.
The requirements for a 1D tight-binding chain to have eigenstates of paired holes (or some suitable paired excitation) are that (i) there exists 2 flavors or species of excitation, (ii) the exchange couplings change the flavor when the particles hop, and (iii) there is a hard-core constraint preventing double occupation on a site.
The Fermi-Hubbard tight-binding chain with onsite repulsion $U$ can, in the hard-core interaction limit $U\to\infty$, fulfill these requirements, albeit in a nonstandard basis.
A recent work by Mamaev et al. \cite{mamaev_Quantum_2019} introduced a proposal of a Fermi-Hubbard model with a staggered laser drive that flips spin on each site.
This laser drive induces an effective spin-orbit coupling that causes spin-flip (flavor-change) hopping in an appropriate excitation basis. 
In the hard-core repulsion limit, hole-paired states emerge as eigenstates of the Hamiltonian.

The model introduced in Ref.~\cite{mamaev_Quantum_2019} is of a 1D Fermi-Hubbard lattice with an added spin-orbit coupling (SOC) laser drive, $\hat{H} = \hat{H}_0 + \hat{H}_{\rm SOC}$, where
\begin{align}
    \hat{H}_0 &= J \sum_{j,\sigma} \left(\hat{c}^\dagger_{j,\sigma}\hat{c}_{j+1,\sigma} + {\rm h.c.}\right) + \frac{U}{2}\sum_j \hat{n}_{j,\uparrow}\hat{n}_{j,\downarrow}, \\ 
    \hat{H}_{\rm SOC} &= \frac{\Omega}{2}\sum_j (-1)^j (\hat{c}_{j,\uparrow}^\dagger\hat{c}_{j,\downarrow} + {\rm h.c.}).
\end{align}
Here $J$ is the hopping rate (taken to be uniform for simplicity, but can be generalized to non-uniform $J$, like the spin chains), $U$ is the Hubbard potential, $\Omega$ is the laser Rabi drive strength, and $\hat{n}_{j,\sigma} = \hat{c}_{j,\sigma}^\dagger\hat{c}_{j,\sigma}$.
As in Ref.~\cite{mamaev_Quantum_2019}, we define a new set of fermion operators
\begin{align}
    \hat{a}_{j,\uparrow} &= \frac{1}{\sqrt{2}}\left( \hat{c}_{j,\uparrow} + (-1)^j\hat{c}_{j,\downarrow} \right), \\ 
    \hat{a}_{j,\downarrow} &= \frac{1}{\sqrt{2}}\left( \hat{c}_{j,\uparrow} - (-1)^j\hat{c}_{j,\downarrow} \right),
\end{align}
which obey the canonical anticommutation relations $\{\hat{a}_{j,\sigma},\hat{a}_{k,\tau}\} = \delta_{jk}\delta_{\sigma \tau}$.
In this basis, the Hamiltonian is
\begin{align}
    \hat{H}&=J\sum_{j,\sigma}\left(\hat{a}_{j,\sigma}^{\dagger}\hat{a}_{j+1,\bar{\sigma}}+\mathrm{h.c.}\right)+\frac{U}{2}\sum_{j}\hat{n}_{j,\uparrow}\hat{n}_{j,\downarrow} \\
    &\qquad+\frac{\Omega}{2}\sum_{j}\Big(\hat{n}_{j,\uparrow}-\hat{n}_{j,\downarrow}\Big), \nonumber
\end{align}
where $\bar{\sigma}$ denotes the spin-flip of $\sigma$ and now $\hat{n}_{j,\sigma} = \hat{a}_{j,\sigma}^\dagger\hat{a}_{j,\sigma}$.
We thus have the desired spin-flip hopping.
Also note the energy splitting $\Omega$ between spin-up and spin-down particles, thus making this a good basis with respect to the SOC laser drive.

Whereas Ref.~\cite{mamaev_Quantum_2019} explores the physics in the limit $\Omega=U\to\infty$, here we wish to consider a slightly different limit: $U\to\infty$ with $\Omega,J<\infty$.
This effects the on-site hard-core repulsion between opposite spins, thus we have the effective model
\begin{align}
    \hat{H}_\infty &= \frac{\Omega}{2}\sum_{j}\Big(\hat{n}_{j,\uparrow}-\hat{n}_{j,\downarrow}\Big)\\
    &+J\prod_{k}\Big(1-\hat{n}_{k,\uparrow}\hat{n}_{k,\downarrow}\Big)\sum_{j\sigma}\left(\hat{a}_{j,\sigma}^{\dagger}\hat{a}_{j+1,\overline{\sigma}}+\mathrm{h.c.}\right). \nonumber
\end{align}
With the hard-core repulsion, the $N$-particle ferromagnetic states $|\Xi_\uparrow\rangle = |{\uparrow}{\uparrow}{\uparrow}\cdots\rangle$ and $|\Xi_\downarrow\rangle = |{\downarrow}{\downarrow}{\downarrow}\cdots\rangle$ are eigenstates of $\hat{H}_\infty$, with energies $+N\frac{\Omega}{2}$ and $-N\frac{\Omega}{2}$, respectively.
We now introduce the fermionic hole-pairing operators (one for each spin)
\begin{align}
    \hat{Q}_\sigma = \frac{1}{\sqrt{N}}\sum_j (-1)^j \hat{a}_{j,\sigma}\hat{a}_{j+1,\sigma},
\end{align}
that create pairs of adjacent holes with staggered phases, in perfect analogy with Eq.~\eqref{eq:Qpair}.
We also denote by $\hat{\mathcal{P}}_\sigma$ the projector into each spin subspace, in analogy with Eq.~\eqref{appeq:proj}.

Just as with the dimer spin chain case, we find that the hole-pairing operators here commute with the tight-binding Hamiltonian when restricted to the appropriate subspace, but the total energy changes due to the removal of two particles:
\begin{align}
    [\hat{Q}_\uparrow,\hat{H}_\infty]\hat{\mathcal{P}}_\uparrow = -\Omega\hat{Q}_\uparrow,\quad[\hat{Q}_\downarrow,\hat{H}_\infty]\hat{\mathcal{P}}_\downarrow = +\Omega\hat{Q}_\downarrow.
\end{align}
Thus, when acting on their respective $|\Xi_\sigma\rangle$, powers of $\hat{Q}_\sigma$ produce towers of Hamiltonian eigenstates with energies ranging from $\pm N\frac{\Omega}{2}$ to either $0$ or $\pm \frac{\Omega}{2}$ for either even-length or odd-length chains, respectively.
Notice that due to the SOC energy splitting, $\hat{Q}_\sigma$ raises or lowers the eigenstate energy by a multiple of the SOC drive strength $\Omega/2$, furthering the analogy with $\eta$-pairing \cite{yang_eta_1989,zhang_Pseudospin_1990}, for which $\eta$-paired states have energy splitting that are a multiple of the Hubbard potential $U$.

%%%%%%%%%%%%%%%%%%%%%%%%%%%%%%%%%%%%%%%%%%%%%%%%%%%%%%%%%%%%%%%%%%%%%%
%%%%%%%%%%%%%%%%%%%%%%%%%%%%%%%%%%%%%%%%%%%%%%%%%%%%%%%%%%%%%%%%%%%%%%

\section{Existence proof for the pure steady state}
\label{app:steady-state-proof}

Here, we will rigorously prove that $|\psi_Q\rangle$, Eq.~\eqref{eq:steady-state}, is indeed a pure steady state of Eq.~\eqref{eq:qme}.
For $|\psi_Q\rangle$ to be a pure steady state, it is sufficient for it to be an eigenstate of the Hamiltonian and a dark state of the dissipation \cite{kraus_Preparation_2008}.
The latter property is satisfied as $\hat{c}|\bullet_1\rangle = \hat{c}|\circ_1\rangle = 0$ and $|\psi_Q\rangle$ has only those components on site 1.
It remains to show that $|\psi_Q\rangle$ is an eigenstate of $\hat{H}$, which we do by way of a variational ansatz.

We consider a two-parameter variational ansatz $|\psi^\prime[\alpha,\beta]\rangle$ and show that it is an exact eigenstate for a set of uniquely determined variational parameters $\alpha$, $\beta$.
First note that for $N=1$, the steady state (cf. Eq.~\eqref{eq:steady-state-2qb}) is
\begin{align}
    |\psi_1\rangle = \left(1+\frac{\Gamma}{\Omega}\hat{\tau}_1 \right)|\bullet\rangle. 
    \label{eq:psi1-dimer-rep}
\end{align}
This is a zero-energy eigenstate of the boundary Hamiltonian 
\begin{align}
    \hat{H}_1 \equiv \hat{H}_{\rm drive}+\hat{H}_{\rm diss}
\end{align}
(cf.~Eqs.~\eqref{appeq:Hdrive-dimer} and \eqref{appeq:Hdiss-dimer}).
Thus we construct a variational ansatz which includes this wavefunction in the $N=1$ case, but we replace $\Gamma/\Omega$ with a variational parameter $\beta$ for generality.
This correction to $|\psi_\infty\rangle$ cannot be enough for $N>1$ because the hole on site 1 will be delocalized throughout the chain in order to have an eigenstate of $\hat{H}_{\rm XX}$.
One might expect that a linear combination of zero energy $\hat{H}_{\rm XX}$ eigenstates with all possible number of hole pairs is needed.
A ``hole-pair condensate'' i.e., the exponential of $\hat{Q}$ acting on $|\psi_\infty\rangle$, is a simple way to achieve that, thus we make the ansatz
\begin{align}
    |\psi^\prime[\alpha,\beta]\rangle = (1+\beta\hat{\tau}_1)e^{\alpha \hat{Q}}|\psi_\infty\rangle.
\end{align}

Evaluating $\hat{H}|\psi^\prime[\alpha,\beta]\rangle$, there are three nonzero terms
\begin{align}
    \hat{H}|\psi^\prime[\alpha,\beta]\rangle = \left[\hat{H}_{1} + \beta \hat{H}_{1}\hat{\tau}_1 + \beta[\hat{H}_{\rm XX},\hat{\tau}_1] \right] e^{\alpha\hat{Q}}|\psi_\infty\rangle,
\end{align}
as $\hat{H}_{\rm XX}e^{\alpha\hat{Q}}|\psi_\infty\rangle = 0$.
In what follows, we use the fact that $e^{\alpha\hat{Q}}|\psi_\infty\rangle \in \mathcal{H}_s$ (cf.~Eq.~\eqref{appeq:Hs}) contains no $|{\sqr}\rangle$ or doublon states $|11\rangle$.
Thus we may always write $e^{\alpha\hat{Q}}|\psi_\infty\rangle = \proj e^{\alpha\hat{Q}}|\psi_\infty\rangle$ (cf.~Eq.~\eqref{appeq:proj}).
Evaluating the boundary Hamiltonian restricted to the the steady state subspace yields
\begin{align}
    \hat{H}_{\rm drive}\proj &= \frac{\Omega}{2}\hat{\lambda}_1^\dagger - \frac{\Delta}{2}\hat{\lambda}_1^\dagger\hat{\tau}_1, \\
    \hat{H}_{\rm diss}\proj &= \frac{1}{4}i\nu\gamma \hat{\lambda}_1^\dagger\hat{\tau}_1.
\end{align}
So we have $\hat{H}_1\proj = \frac{1}{2}\Omega\hat{\lambda}_1^\dagger - \frac{1}{2}\Gamma \hat{\lambda}_1^\dagger\hat{\tau}_1$, where $\Gamma$ is defined in Eq.~\eqref{eq:Gamma}.
Likewise, the commutator is given by $[\hat{H}_{\rm XX},\hat{\tau}_{1}]\proj=\frac{1}{4}J_{1}\hat{\lambda}_{1}^{\dagger}\hat{\tau}_{1}\hat{\tau}_{2}$. 
Thus we have
\begin{align}
    \hat{H}&|\psi^\prime[\alpha,\beta]\rangle = \\
    &\left[\frac{1}{2}\Omega\hat{\lambda}_{1}^{\dagger} + \beta\frac{1}{4}J_{1}\hat{\lambda}_{1}^{\dagger}\hat{\tau}_{1}\hat{\tau}_{2} +\frac{1}{2}\big(\beta\Omega-\Gamma\big) \hat{\lambda}_{1}^{\dagger}\hat{\tau}_{1} \right] e^{\alpha\hat{Q}}|\psi_\infty\rangle.\nonumber
\end{align}
Commuting everything past the exponential $e^{\alpha\hat{Q}}$, one can readily show that $[\hat{\lambda}^\dagger_1\hat{\tau}_1,\hat{Q}]\proj=[\hat{\lambda}^\dagger_1\hat{\tau}_1\hat{\tau}_2,\hat{Q}]\proj=0$.
The only non-zero commutator is
\begin{align}
    [\hat{\lambda}^\dagger_1,e^{\alpha \hat{Q}}]\proj = -\alpha (J_1/2\Jbar\sqrt{N})e^{\alpha\hat{Q}}\hat{\lambda}^\dagger_1\hat{\tau}_1\hat{\tau}_2.
\end{align}
This commutator is evaluated by first noting that $[\hat{\lambda}^\dagger_1,\hat{Q}]\proj = -(J_1/2\Jbar\sqrt{N})\hat{\lambda}^\dagger_1\hat{\tau}_1\hat{\tau}_2$ and $[\hat{\lambda}^\dagger_1\hat{\tau}_1\hat{\tau}_2,\hat{Q}]\proj = 0$.
Thus $[[\hat{\lambda}^\dagger_1,\hat{Q}],\hat{Q}]\proj = 0$.
Then we can use the general results that for two operators $\hat{A}$, $\hat{B}$ satisfying $[[\hat{A},\hat{B}],\hat{B}]=0$, the commutator $[\hat{A},e^{\alpha\hat{B}}] = \alpha [\hat{A},\hat{B}]e^{\alpha\hat{B}}= \alpha e^{\alpha\hat{B}}[\hat{A},\hat{B}]$.
Using the hard-core constraint, we also have $\hat{\lambda}^\dagger_1|\psi_\infty\rangle=0$, hence the action of $\hat{H}$ on the variational ansatz reduces to the following two terms
\begin{align}
    &\hat{H}|\psi^\prime[\alpha,\beta]\rangle = \\
    &e^{\alpha\hat{Q}}\left[\frac{J_{1}}{4}\left(\beta-\alpha\frac{\Omega}{\Jbar\sqrt{N}}\right)\hat{\lambda}_{1}^{\dagger}\hat{\tau}_{1}\hat{\tau}_{2}+\frac{1}{2}\big(\beta\Omega-\Gamma\big)\hat{\lambda}_{1}^{\dagger}\hat{\tau}_{1}\right]|\psi_\infty\rangle.
    \nonumber
\end{align}
Since both terms involve the creation of a $|{\sqr}_1\rangle$, the eigenstate must have eigenvalue 0, and each term must vanish separately as they remove differing numbers of $|\bullet_j\rangle$.
Immediately, we see that letting $\beta = \Gamma/\Omega$ (as predicted by Eq.~\eqref{eq:psi1-dimer-rep}) and $\alpha = \sqrt{N}\Gamma \Jbar/\Omega^2$ yields $|\psi_Q\rangle$ as desired.

We conclude this appendix with an important comment on uniqueness.
While we do not have a proof that this is the \emph{unique} steady state of the master equation for any $N>1$, we have reason to expect that it is unique so long as $|\drive|<\infty$.
As noted above, the original $N=1$ problem does have a unique steady state $|\psi_1\rangle$ (cf. Eq.~\eqref{eq:steady-state-2qb}) for any finite driving strength $|\Omega/\Gamma|<\infty$.
Numerical exact diagonalization of the Liouvillian for up to $N=4$ finds Eq.~\eqref{eq:steady-state} to be the unique steady state.
Moreover, there is no obvious symmetry in the problem that would allow for a steady state degeneracy (except in the limit $\Omega\to\infty$, see Sec.~\ref{sec:entanglement}), thus we expect this steady state to be generically unique.

%%%%%%%%%%%%%%%%%%%%%%%%%%%%%%%%%%%%%%%%%%%%%%%%%%%%%%%%%%%%%%%%%%%%%%
%%%%%%%%%%%%%%%%%%%%%%%%%%%%%%%%%%%%%%%%%%%%%%%%%%%%%%%%%%%%%%%%%%%%%%

\section{Recursive steady state construction}
\label{app:recursion-correlation}

\subsection{Steady state recursion relation}

The steady state $|\psi\rangle$ given by Eq.~\eqref{eq:steady-state} can alternatively be constructed recursively in the length of the chains $N$.
While this construction does not lead to further analytic insights into the structure and properties of the steady state beyond what can be obtained from Eq.~\eqref{eq:steady-state}, it does allow for the numerically amenable calculation of expectation values and correlation functions. 
In what follows, it is convenient to define the dimensionless parameters
\begin{align}
    \zeta \equiv \sqrt{\frac{\Gamma}{\Jbar}}, \quad \eta_j \equiv \frac{J_j}{\Jbar}
\end{align}
The two $\zeta$ and $\drive$ (cf. Eq.~\eqref{eq:eff-drive}) are sufficient to describe overall properties of the steady state (i.e., $|\psi\rangle$ can be written using only these two parameters when the $\{J_j\}$ are hidden away in $\hat{Q}$), and the set of $\{\eta_j\}$ encode all intra-chain hopping disorder.

The starting point for the recursive construction is the pair of steady states $|\psi_1\rangle$ and $|\psi_2\rangle$ for the $N=1$ and $N=2$ chains, respectively.
The $N=1$ steady state $|\psi_1\rangle$ (cf. Eq.~\eqref{eq:steady-state-2qb}) is given in terms of $\drive$ and $\zeta$ by
\begin{align}
    |\psi_1\rangle &= \frac{1}{\mathcal{N}_1} \left[ \sqrt{2}\frac{\zeta}{\drive}|\circ_1\rangle + |\bullet_1\rangle \right],\\
    \mathcal{N}_1^2 &= \frac{2|\zeta|^2}{|\drive|^2} + 1,
\end{align}
where $\mathcal{N}_1$ is the normalization.
For the later use in computing correlation functions, it is crucial that the states be normalized.
The $N=2$ steady state is
\begin{align}
    |\psi_2\rangle &= \frac{1}{\mathcal{N}_2} \left[ \frac{\eta_1}{\mathcal{N}_1\drive^2}|{\circ}_1{\circ}_2\rangle - |\psi_1\rangle\otimes|{\bullet}_2\rangle \right],\\
    \mathcal{N}_2^2 &= \frac{\eta_1^2}{\mathcal{N}_1^2 |\drive|^4} + 1.
\end{align}
By construction this is a dark state of the dissipation, and one can readily verify that it is a zero-energy eigenstate of the Hamiltonian $\hat{H}_2 = \hat{H}_1 + \hat{H}_{{\rm XX},2}$.

For all $n\geq 3$, the steady state is given recursively by
\begin{align}
    |\psi_n\rangle &= \frac{1}{\mathcal{N}_{n}}\left[ \frac{\eta_{n-1}}{\mathcal{N}_{n-1}\drive^2}|\psi_{n-2}\rangle\otimes|{\circ}_{n-1}{\circ}_{n} \rangle\right.\label{appeq:steady-state-rec}\\
    &\qquad\qquad~~\left.+ (-1)^{\lfloor n/2\rfloor} |\psi_{n-1}\rangle\otimes|\bullet_n\rangle \right],\nonumber\\
    \mathcal{N}_n^2 &= \frac{\eta_{n-1}^2}{\mathcal{N}_{n-1}^2 |\drive|^4} + 1. \label{appeq:ss-norm-rec}
\end{align}
One readily verifies that this is an eigenstate of the Hamiltonian $\hat{H}_n = \hat{H}_1 + \hat{H}_{{\rm XX},n}$ using the induction hypotheses $\hat{H}_{n-1}|\psi_{n-1}\rangle = \hat{H}_{n-2}|\psi_{n-2}\rangle = 0$.
The hypotheses hold for $|\psi_1\rangle$ and $|\psi_2\rangle$, thus completing the inductive proof that $\hat{H}_n|\psi_n\rangle = 0$.
Notice that the pairing of holes in the steady state appears clearly in the recursion relation.
The strong $\drive$ limit $|\psi\rangle \to |\bullet\bullet\bullet\cdots\rangle$ is also evident by neglecting terms at least $\mathcal{O}(1/\drive)$ in small $1/\drive\ll 1$.

\subsection{Correlation functions}
\label{app:correlation-funcs}

The recursive construction of $|\psi\rangle$ provides a convenient way to numerically compute correlation functions.
Using Eq.~\eqref{eq:steady-state} directly presents some analytic challenges that are avoided in the recursion.
For the sake of clarity, we focus here on the dimer particle number $\hat{n}_j \equiv \frac{1}{2}\hat{\tau}_j^\dagger\hat{\tau}_j$.
Since the recursion relation for $|\psi_n\rangle$ is in the chain length $n$, it is necessary to denote the chain length for which expectation values are taken, e.g.
\begin{align}
    \langle \hat{n}_j \rangle_n = \frac{1}{2}\langle\psi_n|\hat{\tau}_j^\dagger\hat{\tau}_j|\psi_n\rangle.
\end{align}
Here the state is normalized by construction, $\langle\psi_n|\psi_n\rangle=1$.
Using the recursion relation Eq.~\eqref{appeq:steady-state-rec}, we expand $\langle \hat{n}_j \rangle_n$ in terms of the expectation values evaluated for $n-1$ and $n-2$ length chains:
\begin{align}
    \langle \hat{n}_j \rangle_n = \frac{\eta_{n-1}^2}{\mathcal{N}_n^2\mathcal{N}_{n-1}^2|\drive|^4}\langle\hat{n}_j\rangle_{n-2} + \frac{1}{\mathcal{N}_n^2}\langle\hat{n}_{j}\rangle_{n-1}, \label{appeq:recursive-expectation}
\end{align}
where $\mathcal{N}_k$ are the normalization factors given by Eq.~\eqref{appeq:ss-norm-rec}.

This expression tells us that the expectation value $\langle \hat{n}_j \rangle_n$, evaluated for a chain of length $n$, is given in terms of the expectation value evaluated on chains of length $n-1$ and $n-2$, which are similarly given by the recursive expression Eq.~\eqref{appeq:recursive-expectation}.
This recursion terminates at the expectation value $\langle\hat{n}_j\rangle_j$, i.e. for a length $n=j$ chain.
The termination $\langle \hat{n}_j \rangle_j$ is readily evaluated using Eq.~\eqref{appeq:steady-state-rec} directly and is
\begin{align}
    \langle \hat{n}_j \rangle_j = \langle\psi_j|\hat{n}_j|\psi_j\rangle = \frac{1}{\mathcal{N}_j^2}.
\end{align}
Any expectation value or correlation function can be evaluated in a similar way.

%%%%%%%%%%%%%%%%%%%%%%%%%%%%%%%%%%%%%%%%%%%%%%%%%%%%%%%%%%%%%%%%%%%%%%
%%%%%%%%%%%%%%%%%%%%%%%%%%%%%%%%%%%%%%%%%%%%%%%%%%%%%%%%%%%%%%%%%%%%%%

\section{Restricted spectrum generating algebra and relation to scarring}
\label{app:rsga-scar}

Here we briefly demonstrate that $\hat{Q}$ and $|\psi_\infty\rangle$ (cf.~Eq.~\eqref{eq:steady-state-strong-drive}) constitute a restricted spectrum generating algebra (RSGA), and show that these states have area-law entanglement entropy across a particular bipartition, and thus are scar states of a simple ladder model.

\subsection{RSGA in a nonintegrable model}

In Ref.~\cite{moudgalya_etapairing_2020}, a restricted spectrum generating algebra of order $M$ for a system with Hamiltonian $\hat H$ is defined by an operator $\hat{\eta}^\dagger$ and a state $|\psi_0\rangle$ satisfying: (i) $\hat{H}|\psi_0\rangle = E_0|\psi_0\rangle$, (ii) $\hat{H}_1|\psi\rangle = \mathcal{E}\hat\eta^\dagger|\psi_0\rangle$, $\hat{H}_n|\psi_0\rangle=0~\forall~n, 2\leq n\leq M$, and (iv) $\hat{H}_n \neq 0,~n\leq M$, $\hat{H}_n = 0,~n=M+1$. Here $\hat{H}_n = [\hat{H}_{n-1},\hat\eta^\dagger]$ are successive commutators with $\hat\eta^\dagger$, and $\hat {H}_0=\hat{H}$ is the system Hamiltonian. Then the states $(\hat\eta^\dagger)^n|\psi_0\rangle$ are eigenstates of $\hat H$ with equally spaced energies $E_0 + n \mathcal{E}$, for all $n$ for which $(\hat\eta^\dagger)^n|\psi_0\rangle \neq 0$.

We consider a simple nonintegrable ladder model, specifically the model discussed in Ref.~\cite{znidaric_Coexistence_2013} of two XX chains with an $XXZ$ interaction on each rung (to which we add a term $\hat H_\delta$):
\begin{align}
    \hat H &= \hat H_{\rm XX} + \hat H_{\rm rung} + \hat H_\delta, \label{appeq:ladder-Hamiltonian}\\
    \hat H_{\rm rung} &= \sum_{j=1}^{L} \left[g\left( \hat\sigma^x_{A,j}\hat\sigma^x_{B,j} + \hat\sigma^y_{A,j}\hat\sigma^y_{B,j}\right) + \mu \hat\sigma^z_{A,j}\hat\sigma^z_{B,j}\right],\nonumber\\
    \hat H_\delta &= \delta\sum_{i=1}^{L/2} \left( \hat\sigma_{A,2i}^+\hat\sigma_{B,2i}^- + {\rm h.c.} \right).
\end{align}
This model is nonintegrable (even for $\mu = 0$), and it conserves total $Z$-axis magnetization   \cite{znidaric_Coexistence_2013},
\begin{align}
    \hat M = \sum_j \hat\sigma_{A,j}^z+\hat\sigma_{B,j}^z.
\end{align}
The rung terms split the $|S\rangle$ and $|T\rangle$ states: $\hat H_{\rm rung}|S\rangle = -2g|S\rangle$, $\hat H_{\rm rung}|T\rangle = +2g |T\rangle$, and acts as a chemical potential by giving an energy cost to adding either holes or particles: $\hat H_{\rm rung}|\circ\rangle = \mu|\circ\rangle$, $\hat H_{\rm rung}|(11)\rangle = \mu|(11)\rangle$.
The added $\hat H_\delta$ splits the two reference states $|\psi_\infty\rangle = |STST\cdots\rangle$ and $|\tilde\psi_\infty\rangle=|TSTS\cdots \rangle$ via $\hat H_\delta|S_{2i}\rangle = -2\delta |S_{2i}\rangle$ and $\hat H_\delta|T_{2i}\rangle = 2\delta |T_{2i}\rangle$, to lift some degeneracies between the hole-paired states and other eigenstates of the system.

It is straightforward to see that the $\hat Q$ operator constructed in App.~\ref{app:hole-pairing} and the reference state $|\psi_\infty\rangle$ (cf.~Eq.~\eqref{eq:steady-state-strong-drive}) form a RSGA of order 2 under this Hamiltonian. In particular, we have
\begin{align}
    \hat H|\psi_\infty\rangle &= E_0|\psi_\infty\rangle,~~E_0 = \begin{cases}
        0 & L~{\rm even}\\
        -g & L~{\rm odd}
    \end{cases}, \label{appeq:ladder-rsga} \\
    \hat H_1|\psi_\infty\rangle &= 2\mu\hat Q|\psi_\infty\rangle,\nonumber\\
    \hat H_2 |\psi_\infty\rangle &= 0, ~~\hat H_3=0, \nonumber
\end{align}
where $\hat H_n = [\hat H_{n-1},\hat Q]$ with $\hat H_0 = \hat H$.
Eq.~\eqref{appeq:ladder-rsga} implies that the set of hole-paired states
\begin{align}
    |\psi_n\rangle &= \hat Q^n |\psi_\infty\rangle, \label{Q-states} \\
    \hat H |\psi_n\rangle &= (E_0 + 2n\mu)|\psi_n\rangle,
\end{align}
are a tower of exact eigenstates of the nonintegrable ladder, with equal energy spacing $E_n - E_{n-1} = 2n\mu$, for all $n<L/2$.

Similar calculations show that $\hat Q^\dagger$ acting on $|\psi_\infty\rangle$ is also a RSGA of order 2, but this time creating particle (i.e., $|(11)_j\rangle$) pairs. Similarly, if we define a $\tilde Q$ which is identical to $\hat Q$ but with the replacement $\hat\tau_j\hat\tau_{j+1} \mapsto \hat\lambda_j\hat\lambda_{j+1}$, and a new reference state $|\tilde\psi_\infty\rangle = |TSTSTS\cdots\rangle$, we also find that $\tilde Q$ and $\tilde Q^\dagger$ with $|\tilde\psi_\infty\rangle$ form order 2 RSGAs, creating hole and particle pairs, respectively.

\subsection{Scarring in a nonintegrable ladder model}

For the hole-paired states to be true quantum scars in the ladder model, it is not enough that they are generated from a restricted spectrum generating algebra.
A key hallmark of quantum scarring is the existence of eigenstates of a nonintegrable system which are highly anomalous in some observable quantities when compared to other eigenstates of similar energy and thus violate strong eigenstate thermalization hypothesis (ETH), which one would otherwise naively expect to hold \cite{moudgalya_Quantum_2022}. 
One common quantity to compare is the entanglement entropy (EE) across some biparition of the quantum system. Generically, eigenstates of nonintegrable systems have volume law EE across a given bipartition; in a 1D system we thus expect $S\sim L$ for a generic eigenstate.

In the ladder model, we consider the bipartition defined by cutting the system lengthwise at site $L/2$ (letting the partitions $\alpha,~\beta$ be the left, right halves of the chain, respectively). It is clear that the reference states $|\psi_\infty\rangle,~|\tilde\psi_\infty\rangle$ have zero EE across the bipartition, as they are tensor products of a Bell state on each dimer site. Furthermore, as we show rigorously below, the EE of hole-paired states follows an area law in the limit $L\to\infty$: $S_\alpha \sim L^0$, thus these eigenstates are quantum scars.

\begin{figure}[t]
    \centering
    \includegraphics[width=0.98\columnwidth]{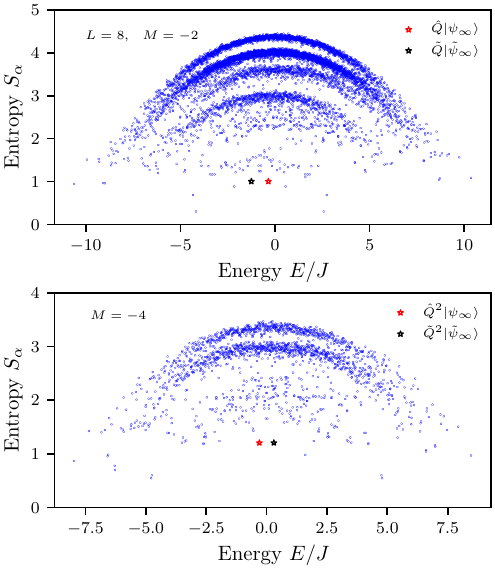}
     \caption{
        \textbf{Hole-paired states as scars in the double XX chain ladder model.} 
        The entanglement entropy of the ladder model eigenstates (cf.~Eq.~\eqref{appeq:ladder-Hamiltonian}) of the $\alpha$ partition (i.e., sites 1 to $L/2$) is plotted vs. the eigenstate energy. Here we consider a length $L=8$ ladder (16 total spins) and plot the eigenstates in the $M=-2$ magnetization sector (i.e., 2-hole sector) and the $M=-4$ (4-hole) sector. 
        The hole-paired states are denoted with stars; in each magnetization sector there are two states corresponding to hole-pairs created in each reference state $|\psi_\infty\rangle$ and $|\tilde\psi_\infty\rangle$.
        For both plots the parameters are $\mu/J = 0.2$, $g/J = 0.5$, and $\delta/g = 0.3$.
    }
     \label{fig:scar}
 \end{figure}

To demonstrate the anomolously low EE of the hole-paired states, we perform numerical exact diagonalization of Eq.~\eqref{appeq:ladder-Hamiltonian} for a length $L=8$ chain (i.e., 16 total spins) using QuSpin \cite{weinberg_QuSpin_2017}. For two fixed magnetization sectors, $M=-2,-4$, we plot the EE of each eigenstate across the central bond vs. eigenstate energy in Fig.~\ref{fig:scar}.
We find that, as is typical of nonintegrable systems, the typical energy eigenstate near the center of the band has high EE (which we expect to be $\sim L$).
The hole-paired states, identified by stars in the figure, are states in the middle of the eigenspectrum and clearly have anomalously low EE.

\subsection{Area law entanglement entropy of hole-paired states}

For completeness, we rigorously show that the entanglement entropy of the hole-paired states across the central cut at $L/2$ of the length $L$ chain follows an area law, scaling as $S(\hat\rho_{\alpha,n}) \sim L^0$, where $\hat\rho_{\alpha,n} = {\rm tr}_\beta [\hat{Q}^n|\psi_\infty\rangle\langle\psi_\infty|\hat{Q}^{\dagger n}]$. Here we denote the partitions by $\alpha,~\beta$; the $\alpha$ partition contains the first $L/2$ qubits of each chain $A$ and $B$.
First, we split the $\hat Q$ operator into three terms:
\begin{align}
    \hat{Q} = \hat{Q}_\alpha + \hat{Q}_\beta + \hat{Q}_{L/2},
\end{align}
where the first term acts only on the $\alpha$ partition (sites 1 through $L/2$) and the second on the $\beta$ partition and the last term $\hat Q_{L/2} \propto \hat\tau_{L/2}\hat\tau_{L/2+1}$ places a hole pair across the central cut.
Now using the fact that the $Q$ states are orthogonal,
\begin{align}
    \langle\psi_\infty|\hat Q^{\dagger m}\hat Q^n|\psi_\infty\rangle \propto \delta_{nm},
\end{align}
specifically that this relation holds term-by-term, and that $\hat Q_{L/2}^2 =0$, we can write a generic $Q$ state as
\begin{align}
    \hat Q^n|\psi_\infty\rangle &= (\hat Q_\alpha+\hat Q_\beta)^n|\psi_\infty\rangle_\alpha|\psi_\infty\rangle_\beta \\
    &+ (\hat Q_\alpha+\hat Q_\beta)^{n-1}\hat Q_{L/2}|\psi_\infty\rangle. \nonumber
\end{align}
Because $\hat Q_{\alpha/\beta}$ only act on their respective halves of the chain, the terms $\propto (\hat Q_\alpha +\hat Q_\beta)^n|\psi_\infty\rangle$ can always be written $\hat Q_\alpha^k|\psi_\infty\rangle_\alpha\otimes\hat Q_\beta^{n-k}|\psi_\infty\rangle_\beta$. Similarly for the $\propto\hat Q_{L/2}$ terms, the only entangling operator is $\hat Q_{L/2}$, with all powers of $\hat Q_{\alpha/\beta}$ only acting on their halves of the chain.
Thus we can efficiently perform the partial trace over the $\beta$ partition of the density matrix.
In particular, terms with unique powers of $\hat Q_\beta$ and $\hat Q_{L/2}$ are orthogonal, thus the partial trace over the $\beta$ partition is
\begin{align}
    &\hat{\rho}_{\alpha,n} = {\rm tr}_\beta \left[\hat Q^n|\psi_\infty\rangle\langle\psi_\infty|\hat Q^{\dagger n}\right] \\
    &~~~~~~=\sum_{k=0}^{n} \binom{n}{k}^2 {||}\hat Q_\beta^{n-k}|\psi_\infty\rangle_\beta{||}^2 \hat Q_\alpha^k|\psi_\infty\rangle_\alpha\langle\psi_\infty|\hat Q^{\dagger k} \nonumber \\
    &+\sum_{k=0}^{n-1} \binom{n-1}{k}^2{||}\hat Q_\beta^{n-1-k}|{\circ}\psi_\infty\rangle_\beta{||}^2 \hat Q_\alpha^k|\psi_\infty{\circ}\rangle_\alpha\langle\psi_\infty{\circ}|\hat Q^{\dagger k}. \nonumber
\end{align}
Here $|{\circ}\psi_\infty\rangle_\beta$ is the $\beta$ partition of $\hat Q_{L/2}|\psi_\infty\rangle$, which has a hole on site $L/2+1$.

For a given fixed $n$, when we take the limit $L\to\infty$ we have ${||}\hat Q_{\alpha/\beta}^{n}|\psi_\infty\rangle_\beta{||}^2\sim (L/2)^n$, as each half of the chain has $\approx L/2$ places where each hole-pair can be placed.
Thus ${\rm tr}[\hat\rho_{\alpha,n}] \sim (L/2)^n\binom{2n}{n} + \mathcal{O}(L^{n-1})$.
Denoting the probabilities of terms where a hole-pair does not span the central cut by $p_k$, and denoting the probabilities of terms with a hole-pair spanning the cut by $\tilde p_k$, we find
\begin{align}
    p_k \sim \binom{n}{k}^2\bigg/\binom{2n}{n},~~\tilde p_k \sim L^{-1} \binom{n-1}{k}^2\bigg/\binom{2n}{n}.
\end{align}
We can neglect the $\tilde p_k$ for $L\to\infty$ (keeping $n$ fixed), so the entanglement entropy of a hole-paired state is area-law:
\begin{align}
    S(\hat\rho_{\alpha,n}) \approx -\sum_k p_k\ln p_k \sim L^0.
\end{align}
Note that this argument only holds for a $2n$-hole state when taking $L\to\infty$ with $n$ fixed.

%%%%%%%%%%%%%%%%%%%%%%%%%%%%%%%%%%%%%%%%%%%%%%%%%%%%%%%%%%%%%%%%%%%%%%
%%%%%%%%%%%%%%%%%%%%%%%%%%%%%%%%%%%%%%%%%%%%%%%%%%%%%%%%%%%%%%%%%%%%%%

\section{Hole-pair correlation function}
\label{app:hole-correlation}

Here we derive the correlation function Eq.~\eqref{eq:hole-density-correlation-single} and discuss its nonstandard normalization. It is natural to measure hole correlations using the dimer operators $\hat \tau_j$ defined in Eq.~\eqref{appeq:tau-op} as $\hat \tau_j^z = [\hat\tau_j^\dagger,\hat\tau_j]$ has the matrix elements $\langle\circ_j|\hat\tau_j^z|\circ_j\rangle = -2$ and $\langle\bullet_j|\hat\tau_j^z|\bullet_j\rangle = 0$ (and zero off-diagonal matrix elements) within the steady state subspace $\{|\circ_j\rangle,|\bullet_j\rangle\}$. Thus, we define the hole-pair correlation function as 
\begin{align}
    C_{zz}(j,k) = \frac{\langle \hat\tau_j^z\hat\tau_k^z\rangle - \langle \hat\tau_j^z\rangle\langle\hat\tau_k^z\rangle}{\sqrt{\big(\langle \hat\tau_j^{z}\tau_j^{z}\rangle - \langle \hat\tau_j^z\rangle^2\big)\big(\langle \hat\tau_k^{z}\hat\tau_k^{z}\rangle - \langle \hat\tau_k^z\rangle^2\big)}},
\end{align}
which is a standard connected correlation function normalized by the onsite fluctuations. To relate this to measurable quantities in the single $A$ chain alone, we first note that $\hat\tau_j^z = \hat\sigma_{A,j}^z + \hat\sigma_{B,j}^z$, and that within the steady state subspace, $\langle\circ_j|\hat\sigma_{s,j}^z|\circ_j\rangle = -1$ and $\langle\bullet_j|\hat\sigma_{s,j}^z|\bullet_j\rangle = 0$, for $s\in\{A,B\}$. Thus, when evaluating the correlation functions for an arbitrary state within the steady state subspace, we have $\langle\hat\sigma_{A,j}^z\rangle = \langle\hat\sigma_{B,j}^z\rangle$, and for $j\neq k$, it is readily shown that all $\langle\hat\sigma_{s,j}^z\hat\sigma_{s^\prime,k}^z\rangle$ are equivalent for any $s,s^\prime\in\{A,B\}$. Therefore, the connected correlation function for $j\neq k$ can be expressed in $A$-chain observables as
\begin{align}
    \langle \hat\tau_j^z\hat\tau_k^z\rangle - \langle \hat\tau_j^z\rangle\langle\hat\tau_k^z\rangle = 4( \langle \hat\sigma_{A,j}^z\hat\sigma_{A,k}^z\rangle - \langle \hat\sigma_{A,j}^z\rangle\langle\hat\sigma_{A,k}^z\rangle).
\end{align}
For $j=k$, the expectation values $\langle\hat\sigma_{s,j}^z\hat\sigma_{s^\prime,j}^z\rangle$ are not all equivalent. The connected correlation functions can be reduced to
\begin{align}
    \langle \hat\tau_j^z\hat\tau_j^z\rangle - \langle \hat\tau_j^z\rangle^2 = 2 + 2\langle \hat\sigma_{A,j}^z\hat\sigma_{B,j}^z\rangle - 4\langle \hat\sigma_{A,j}^z\rangle^2.
\end{align}
As a final step, we find that within the steady state subspace, the nonzero matrix elements of $\hat\sigma_{A,j}^z\hat\sigma_{B,j}^z$ are $\langle\circ_j|\hat\sigma_{A,j}^z\hat\sigma_{B,j}^z|\circ_j\rangle = 1$ and $\langle\bullet_j|\hat\sigma_{A,j}^z\hat\sigma_{B,j}^z|\bullet_j\rangle = -1$. Thus when restricted to this subspace, the expression $2 + 2\langle \hat\sigma_{A,j}^z\hat\sigma_{B,j}^z\rangle$ has the same matrix elements as $-4\langle\hat\sigma_{A,j}^z\rangle$. While these are not equivalent as operators, we thus have the expectation value equivalence
\begin{align}
    2 + 2\langle \hat\sigma_{A,j}^z\hat\sigma_{B,j}^z\rangle = -4\langle\hat\sigma_{A,j}^z\rangle
\end{align}
in the steady state subspace. Therefore, the correlation function defined above in terms of $\hat\tau_j^z$ is equivalent to Eq.~\eqref{eq:hole-density-correlation-single} when evaluated on the steady state (and any state in the steady state subspace, generally).

%%%%%%%%%%%%%%%%%%%%%%%%%%%%%%%%%%%%%%%%%%%%%%%%%%%%%%%%%%%%%%%%%%%%%%
%%%%%%%%%%%%%%%%%%%%%%%%%%%%%%%%%%%%%%%%%%%%%%%%%%%%%%%%%%%%%%%%%%%%%%

\section{Approximate coherent states of hole-pairs}
\label{app:univ-state-norms}

The analogy we make to bosonic coherent states in Sec.~\ref{sec:hole-pairing} can be made precise for large $N\gg1$.
For the sake of clarity, we consider the limit $\zeta\to 0$ to focus only on the exponential of $\hat{Q}$ in  $|\psi\rangle$ (cf. Eq.~\eqref{eq:ss-rewrite}).
We seek to show that the probability distribution of finding $m$ hole-pairs in the chain is a Poisson distribution when $m\ll N$, hence $|\psi\rangle = e^{\alpha\hat{Q}}|\psi_\infty\rangle$ approximates a coherent state with displacement $\alpha=\sqrt{N}/\drive^2$ from the ``hole vacuum'' $|\psi_\infty\rangle$.

Each power of $\hat{Q}$ acting on $|\psi_\infty\rangle$ adds one more hole pair to the state.
Thus, the probability $p(m)$ for having $m$ hole pairs in the state is
\begin{align}
    p(m) = \frac{1}{Z} \frac{(\sqrt{N}/|\drive|^{2})^{2m}}{(m!)^2} \langle\psi_\infty|\hat{Q}^{\dagger m}\hat{Q}^m|\psi_\infty\rangle,
\end{align}
where $Z$ is an overall normalization.
The crucial step here is to compute the state norms ${||}\hat{Q}^m|\psi_\infty\rangle{||}^2 = \langle\psi_\infty|\hat{Q}^{\dagger m}\hat{Q}^m|\psi_\infty\rangle$.
If the state norms for small $m\ll N$ are ${||}\hat{Q}^m|\psi_\infty\rangle{||}^2 \approx m!$,
then we have the Poisson distribution
\begin{align}
    p(m) \approx e^{-N/|\drive|^{4}}\frac{(N/|\drive|^{4})^{m}}{m!},
\end{align}
and therefore the average number of holes in the state is $\langle \hat{M}\rangle = 2N/|\drive|^4$ (twice the number of hole-pairs).
We thus arrive at the result that for $\langle \hat{M}\rangle \ll N$, the density of holes is \emph{intensive}
\begin{align}
    \bar{m} = \frac{1}{N}\langle\hat{M}\rangle = \frac{2}{|\drive|^4},
\end{align}
and depends only on $\drive$.
It only remains to compute the state norms and verify ${||}\hat{Q}^m|\psi_\infty\rangle{||}^2 \approx m!$.

The $m$-hole-pair state norms ${||}\hat{Q}^m|\psi_\infty\rangle{||}^2$ are found by simply counting the configurations of $m$ hole pairs in a length $N$ chain.
First consider a single hole-pair.
There are $(N-1)/1!$ unique configurations for a single hole-pair in the chain, and each of those configurations has an amplitude $(1!)^2$, thus the state norm for one hole-pair is 
\begin{align}
    {||}\hat{Q}|\psi_\infty\rangle{||}^2 &= \frac{1}{N} \frac{N-1}{1!} (1!)^2 \approx 1!,
\end{align}
with an $\mathcal{O}(1/N)$ correction.
Moving to two hole-pairs, we see that there are still $N-1$ configurations for the first pair, but now due to the hard-core constraint there are only $N-3$ configurations for the second.
Note that here, we neglect the rare configurations in which the first pair limits the second pair to $N-4$ configurations (i.e., when the first pair is 1 site from the boundary).
Thus, there are $\approx(N-1)(N-3)/2!$ unique configurations of two hole-pairs, each with amplitude $(2!)^2$, thus the state norm is
\begin{align}
    {||}\hat{Q}^2|\psi_\infty\rangle{||}^2 &\approx \frac{1}{N^2} \frac{(N-1)(N-3)}{2!} (2!)^2 \approx 2!.
\end{align}
We may proceed in this way for larger $m$, noting that the number of unique configurations for $m$ holes is $\approx N^m/m! + \mathcal{O}(1/N)$.
Therefore, for $m\ll N$ we have ${||}\hat{Q}^m|\psi_\infty\rangle{||}^2 \approx m!$, the desired result.

We pause here to note that this line of argument will certainly break down when $m \sim N$ as the hard-core constraint allows significantly fewer unique configurations of $m$ hole-pairs than $N^m/m!$.
A more careful analysis of the errors suggests that the break down of the analogy may occur for an even tighter constraint $m \sim \sqrt{N}$; nevertheless, direct numerical computation of the hole density using the exact solution shows a remarkable adherence to the estimate $m\approx N/|\drive|^4$ for much higher hole densities than would be reasonably expected from this analysis.
The precise reason for this is not yet understood.

Finally, having established the coherent state analogy in the $|\zeta|\to0$ limit, we consider the general case.
All of the analysis above proceeds in exactly the same way, but now the dissipative site has extra probability of containing a hole due to the $\propto\zeta$ term in Eq.~\eqref{eq:ss-rewrite}.
The state can be written as the sum of a hole coherent state on all sites and a hole coherent state on sites 2 through N with an isolated hole on site 1:
\begin{align}
    |\psi\rangle = e^{\alpha\hat{Q}}|\psi_\infty\rangle + \frac{\sqrt{2}\zeta}{\drive}e^{\alpha\hat{Q}^\prime}|{\circ}_1\psi_\infty^\prime\rangle,
\end{align}
where $\hat{Q}^\prime$ and $|\psi_\infty^\prime\rangle$ denote those objects defined on sites 2 through $N$ and where $\alpha = \sqrt{N}/\drive^2$.
For large $N\gg1$, the state norms are approximately equal: ${||}e^{\alpha\hat{Q}}|\psi_\infty\rangle{||}^2 \approx {||}e^{\alpha\hat{Q}^\prime}|{\circ}_1\psi_\infty^\prime\rangle{||}^2$.
Using this fact and that the term proportional to $\zeta$ has one more hole than the $\zeta$-independent term, we find the number of holes in the state to be
\begin{align}
    \langle\hat{M}\rangle = \frac{2N}{|\drive|^4} + \frac{2|\zeta|^2}{|\drive|^2+2|\zeta|^2}.
\end{align}
The first term is extensive and recovers the universal scaling $\bar{m}= 2/|\drive|^{4}$ of the hole coherent state, and the second term is independent of $N$ and reflects the additional hole density on site 1, which approaches 1 additional hole in the limit $|\zeta|\to\infty$.

%%%%%%%%%%%%%%%%%%%%%%%%%%%%%%%%%%%%%%%%%%%%%%%%%%%%%%%%%%%%%%%%%%%%%%
%%%%%%%%%%%%%%%%%%%%%%%%%%%%%%%%%%%%%%%%%%%%%%%%%%%%%%%%%%%%%%%%%%%%%%

\section{Emergence of charge density waves}
\label{app:CDW-state-norms}

Here we derive the drive strength regime in which the single particle CDWs emerge, i.e. Eqs.~\eqref{eq:cdw-regime-odd} and \eqref{eq:cdw-regime-even}.

\subsection{{\it \textzeta}-independent upper bound}

We first consider the limit $|\zeta|\to0$ in which odd-length chains have CDWs.
We estimate the particle density of even and odd length chains by expanding $|\psi\rangle$ to second order in $|\drive|$:
\begin{align}
    |\psi\rangle\approx\hat{Q}^{\lfloor N/2\rfloor-1}|\psi_\infty\rangle+\frac{\sqrt{N}}{\tilde{\Omega}^{2}}\frac{1}{\lfloor N/2\rfloor}\hat{Q}^{\lfloor N/2\rfloor}|\psi_\infty\rangle.
\end{align}
For even chains, the particle numbers of the two terms are 2 and 0, respectively, while for odd chains the particle numbers are 3 and 1.
Thus the total chain particle numbers are (using subscripts $e$ and $o$ for even- and odd-length chains, respectively):
\begin{align}
    \langle\hat{N}\rangle_e &= \frac{2}{1+\frac{4}{N|\drive|^4}\frac{{||}\hat{Q}^{N/2}|\psi_\infty\rangle{||}^2}{{||}\hat{Q}^{N/2-1}|\psi_\infty\rangle{||}^2}}, \\
    \langle\hat{N}\rangle_o &= \frac{3+\frac{4}{N|\drive|^4}\frac{{||}\hat{Q}^{\lfloor N/2\rfloor}|\psi_\infty\rangle{||}^2}{{||}\hat{Q}^{\lfloor N/2\rfloor-1}|\psi_\infty\rangle{||}^2}}{1+\frac{4}{N|\drive|^4}\frac{{||}\hat{Q}^{\lfloor N/2\rfloor}|\psi_\infty\rangle{||}^2}{{||}\hat{Q}^{\lfloor N/2\rfloor-1}|\psi_\infty\rangle{||}^2}}.
\end{align}
The onset of the CDW scale occurs when these particle numbers start to diverge from each other -- $\langle\hat{N}\rangle_e$ vanishes with decreasing $\drive$ whereas $\langle\hat{N}\rangle_o$ saturates to 1 particle.
We thus need to estimate the ratio of state norms.

We do not need to estimate the absolute magnitude of the state norms, but only their ratios.
Ignoring any boundary effects and assuming that all hole configurations of each state are equally likely, the state norm of $\hat{Q}^{k}|\psi_\infty\rangle$ can be given as $A(k) \times ({\rm \#~configs~of}~k~{\rm hole{-}pairs})$ where $A(k)$ is an amplitude weighting function.
Dropping the floor notation for simplicity, for the nearly empty chains, each configuration of $\hat{Q}^{N/2-1}|\psi_\infty\rangle$ produces up to one configuration of $\hat{Q}^{N/2}|\psi_\infty\rangle$ (neglecting the $\mathcal{O}(1/N)$ rare configurations of three adjacent particles in odd chains) with a permutation factor $N/2$.
Thus, $A(N/2) = \frac{1}{N}(N/2)A(N/2 -1)$, where the extra $1/N$ comes from the normalization of $\hat{Q}$ (cf. Eq.~\eqref{eq:Qpair}).

For even chains, there are $(N/2)(N/2-1)/2$ configurations of $N/2-1$ hole-pairs (i.e., 2 particles), of which $N/2$ produce the vacuum state (i.e., $N/2$ hole-pairs).
Thus, the state norm ratio for even chains is
\begin{align}
    \frac{{||}\hat{Q}^{\lfloor N/2\rfloor}|\psi_\infty\rangle{||}^2_e}{{||}\hat{Q}^{\lfloor N/2\rfloor-1}|\psi_\infty\rangle{||}^2_e} \approx \frac{\frac{1}{N}(N/2)A(N/2-1)(N/2)}{A(N/2-1)(N/2)^2/2} = \frac{2}{N}.
\end{align}
Similarly, for odd chains, there are $(N/2)(N/2+1)(N/2+2)/6$ configurations of $N/2-1$ hole pairs (i.e., 3 particles), of which $(N/2-1)(N/2)/2$ produce configurations of $N/2$ hole pairs (i.e., 1 particle CDW), thus
\begin{align}
    \frac{{||}\hat{Q}^{\lfloor N/2\rfloor}|\psi_\infty\rangle{||}^2_o}{{||}\hat{Q}^{\lfloor N/2\rfloor-1}|\psi_\infty\rangle{||}^2_o} \approx \frac{\frac{1}{N}(N/2)A(N/2-1)(N/2)^2/2}{A(N/2-1)(N/2)^3/6} = \frac{3}{N}.
\end{align}

From these state norms, we immediately obtain the particle numbers
\begin{align}
    \langle \hat{N}\rangle_e &\approx \frac{1}{4}N^2 |\drive|^4, \label{appeq:even-chain-no-cdw-density}\\
    \langle \hat{N}\rangle_o &\approx 1 + \frac{1}{6}N^2|\drive|^4,
\end{align}
which are valid when we assume the drive strength satisfies
\begin{align}
    |\drive| \ll \frac{1}{\sqrt{N}}.
\end{align}
Thus we have the upper bound on $|\drive|$ below which CWDs emerge in odd length chains in the $|\zeta|=0$ limit.

This analysis is essentially identical in the $|\zeta|\to\infty$ limit via the same reasoning used to explain the emergence of single particle states in even-length chains.
Because site 1 \emph{always} has a hole, we may basically neglect it and consider chains of length $N{-}1$ in an effective $|\zeta|=0$ limit, thus obtaining exactly the same CDW emergence scale.
The only difference here is that even chains have CDWs and odd chains do not.

\subsection{{\it \textzeta}-dependent lower bound}

With the upper bound of the CDW regime established, we now turn to the $|\zeta|$-dependent lower bound.
This bound is characterized by the point at which the particle number of chains with CDWs starts to fall below 1.
Note that the CDWs persist below the lower bound, but with a particle number that vanishes as $\sim|\drive|^2$, i.e., there is less than one particle in the chain on average.

Here we focus on the case of odd chains for $|\zeta|\ll1$ -- as above the even chain in the $|\zeta|\gg1$ regime follows analogously.
We expand the odd chain state to first order in $|\drive|$ as
\begin{align}
    |\psi\rangle_o = \hat{Q}^{N/2}|\psi_\infty\rangle + \frac{\zeta}{\drive}\hat{\tau}_1|\psi_\infty\rangle.
\end{align}
These two terms are the CDW and vacuum, respective.
Note that only one configuration of the CDW can be acted upon with $\hat{\tau}_1$ to produce vacuum, hence we immediately obtain the state norm ratio ${||}\hat{Q}^{N/2}|\psi_\infty\rangle{||}^2/{||}\hat{Q}^{N/2}\hat{\tau}_1|\psi_\infty\rangle{||}^2 = N/2$.
Thus, computing the particle number yields
\begin{align}
    \langle\hat{N}\rangle_o = \frac{\frac{1}{2}N|\zeta|^2|\drive|^2}{1+\frac{1}{2}N|\zeta|^2|\drive|^2},
\end{align}
from which we immediately derive the lower bound on the single particle CDW for odd chains
\begin{align}
    |\drive| \gg \frac{|\zeta|}{\sqrt{N}}.
\end{align}
Note that for the even chain, we have the same condition but with the inverse $|\zeta|^{-1}$.
Finally we note that if $|\zeta|\approx 1$, the CDWs of either parity chain will not be particularly pronounced because the vacuum will be reached before the 2 and 3-particle states are fully suppressed.

%%%%%%%%%%%%%%%%%%%%%%%%%%%%%%%%%%%%%%%%%%%%%%%%%%%%%%%%%%%%%%%%%%%%%%
%%%%%%%%%%%%%%%%%%%%%%%%%%%%%%%%%%%%%%%%%%%%%%%%%%%%%%%%%%%%%%%%%%%%%%

\section{Notes on experimental realization in circuit QED}
\label{app:experiment}

Here, we briefly describe the necessary components and setup to realize on a circuit QED platform both the two-chain model (for remote entanglement stabilization) and the single chain model. 
For the remote entanglement scheme, a detailed study of waveguide and qubit losses will be given in a complementary work \cite{irfan_Loss_2024}.

\subsection{Remote entanglement realization}

We consider two chains of $N$ qubits each, with nearest neighbor capacitive coupling (one may use tunable couplings \cite{karamlou_Quantum_2022}), and Rabi drives applied to qubit 1 of each chain:
\begin{align}
    &\hat H =  \label{appeq:two-chain-H}\\
    &\sum_{s,j}\frac{\omega^q_{s,j}}{2}\hat\sigma^z_{s,j} + \sum_{s,j}\frac{J}{2} \hat\sigma^x_{s,j}\hat\sigma^x_{s,j+1} + \sum_s \frac{\Omega}{2}\cos (\omega_{0}t)\hat\sigma^x_{s,1}. \nonumber
\end{align}
Here, $\omega^q_{s,j}$ are the qubit frequencies, $J$ is the hopping rate, $\Omega$ is the Rabi drive strength, and $\omega_0$ is the common Rabi frequency.
We take qubits 2 through $N$ of each chain to be resonant with each other and with the Rabi drives, thus $\omega^q_{s,j>1} = \omega_0$. If the driven qubits of each chain are \emph{not} resonant with the other qubits, they must be oppositely detuned from $\omega_0$: $\Delta \equiv |\omega^q_{s,1}-\omega_0|$. Note that the scheme is perfectly robust to any disorder in the hopping rates along the two chains so we can replace $J$ with $J_j$ in Eq.~\eqref{appeq:two-chain-H}. Moving to a common rotating frame at $\omega_0$ and making a rotating wave approximation, we arrive at the desired Hamiltonian
\begin{align}
    \hat H &= \frac{\Delta}{2}\left(\hat\sigma^z_{A,1} - \hat\sigma^z_{B,1}\right) + \frac{\Omega}{2}\left(\hat\sigma^x_{A,1}+\hat\sigma^x_{B,1}\right) \\
    &+ \frac{1}{2}\sum_{s,j} J_j\left(\hat\sigma_{s,j}^+\hat\sigma_{s,j+1}^- + {\rm h.c.}\right), \nonumber
\end{align}
(cf.~Eqs.~\eqref{eq:Hdrive} and \eqref{eq:Hxx}). Now it remains to engineer the collective dissipation.

In a remote circuit QED realization, the collective dissipation is engineered using a waveguide that links the two remote chains together. In the main text we consider both bidirectional and unidirectional waveguides.
The driven qubits are coupled to the waveguide with rate $\gamma$. To realize the collective dissipation using a bidirectional waveguide, the qubits must be properly positioned along the waveguide a distance $\Delta x = n\lambda_0/2$ apart, i.e., an integer number of half-wavelengths of the drive frequency $\omega_0$ \cite{govia_Stabilizing_2022}.
Precise spacing control of qubits along a waveguide has been demonstrated in state-of-the-art waveguide QED experiments \cite{fang_Generalized_2017,kannan_Ondemand_2023}.
A unidirectional waveguide can be constructed using microwave circulators that couple the output fields of the qubits to only one propagating direction of the waveguide. Here, there is no spacing requirement to get collective loss. Moreover, the nonreciprocity of the waveguide automatically induces the dissipative exchange Hamiltonian $\hat H_{\rm diss}$ (cf.~Eq.~\eqref{eq:Hdiss}) \cite{metelmann_Nonreciprocal_2015}. Combined with the Hamiltonian derived above, we arrive at the desired master equation, Eq.~\eqref{eq:qme}.

\subsection{Qubit dephasing}

Any experimental realization of a scheme must contend with unwanted dissipation, and the remote entanglement stabilization scheme is no exception. We consider a superconducting circuit realization using transmon qubits. 
The dominant source of dissipation affecting this scheme is unwanted qubit dephasing, as dephasing quickly degrades the coherence of the entangled states. 
Other sources of dissipation, namely qubit relaxation and losses in the waveguide, will be discussed in more detail in a complementary work \cite{irfan_Loss_2024}.
Here we provide an estimate for how much transmon dephasing can be tolerated using realistic driving, hopping and dissipation rates. A heuristic for determining how much qubit dephasing can be tolerated is to compare the typical qubit coherence time $T_2$ to the stabilization time $\tau_{\rm rel}$ of the ideal system (i.e., the relaxation time of the slowest Liouvillian eigenmode). When $\tau_{\rm rel} \ll T_2$, the qubit dephasing will have relatively small effects on the steady state, but when $\tau_{\rm rel} \approx T_2$, the performance of the scheme is rapidly degraded.

Based on previous waveguide QED experiments \cite{kannan_Ondemand_2023,vool_Continuous_2016,mirhosseini_Cavity_2019}, engineered collective dissipation rates have been demonstrated ranging from $\gamma/2\pi = 20 \, {\rm MHz}$  to $\gamma/2\pi = 100 \, {\rm MHz}$. 
Here we take a rather conservative set of parameters $\gamma = \Omega = \Bar{J} = 2\pi \times 3\, {\rm MHz}$.
In the dephasing-free system with $3+3$ qubits, these parameters lead to $90\%$ fidelity to the ideal tensor product of Bell states $|STS\rangle$.
We find the stabilization rate numerically to be $\tau_{\rm rel} \approx 156/\gamma = 9\, \mu s$.  This is considerably faster than the typical qubit coherence time, which have been shown to be typically at least $T_2 \sim 20 \,{\rm \mu s}$ \cite{kannan_Ondemand_2023} up to $T_2 \sim 1\, {\rm ms}$\cite{somoroff_Millisecond_2023}. 
Moreover, as discussed in Sec.~\ref{sec:entanglement}, utilizing the natural third level of transmon accelerates the dissipation process by orders of magnitude. For a $3+3$ system with optimal qutrit coupling, here instead we aim for higher Bell state fidelity of $99\%$ with $\gamma = 2\pi \times 3\, {\rm MHz}$, and $ \Omega = \Bar{J} = 2\pi \times 10\, {\rm MHz}$. We numerically find the stabilization time to be $\tau_{\rm rel} \approx 50/\gamma = 2.7\, {\rm \mu s}$. Thus, even the modest parameter values considered here renders the scheme experimentally feasible and robust against unwanted dephasing.

\subsection{The single spin chain realization}

The single chain model (cf.~Eq.~\eqref{eq:single-chain-qme}) is readily implemented in a variety of platforms including circuit QED. In the circuit QED case, the realization of the dynamics is essentially identical to that of the remote entanglement scheme: starting from a 1D chain of capacitively-coupled qubits, we tune all qubits $j\geq 2$ into resonance with a coherent Rabi drive applied on the $j=1$ qubit. The dissipation on the first qubit can be implemented in any number of standard ways including coupling the qubit to a heavily damped photonic mode. The nearest neighbor hopping rates can be arbitrarily disordered and the driving and dissipation rates can be tuned freely.

The single chain non-equilibrium steady state is sensitive to local qubit dephasing like the remote entanglement scheme, but it has two advantages that make it a more forgiving experiment to perform.  First, as we find in Fig.~\ref{fig:qutrit-scheme}, the single chain relaxation time is typically much shorter than the double chain, even for strong driving (in Fig.~\ref{fig:qutrit-scheme}, $\Omega/\gamma = 10$). Second, we find numerically that the relaxation time of the single chain is shortest when $\gamma \sim \Omega \sim \Jbar$, which is precisely the parameter regime in which the correlation effects in the NESS are strongest. In this parameter regime $|\drive|\sim 1$ which is where many of the magnetization correlations are strongest, as we find in Fig.~\ref{fig:hole-pairing-correlations}.
Using a similar set of parameters as in the remote entanglement case, $\gamma = \Omega = \Jbar = 2\pi \times 3 {\rm MHz}$, we find that a 7-qubit chain has a relaxation time of $\tau_{\rm rel} \approx 2.6 {\rm \mu s}$. Even if the relaxation time scales as $N^3$, we expect that somewhat longer chains could still be experimentally feasible with typical transmon $T_2$.

%%%%%%%%%%%%%%%%%%%%%%%%%%%%%%%%%%%%%%%%%%%%%%%%%%%%%%%%%%%%%%%%%%%%%%
%%%%%%%%%%%%%%%%%%%%%%%%%%%%%%%%%%%%%%%%%%%%%%%%%%%%%%%%%%%%%%%%%%%%%%

\section{Stabilization slowdown due to conservation of angular momentum}
\label{app:slowdown}

The necessary and sufficient conditions for the existence of a pure steady state $\hat{\mathcal{L}}|\psi\rangle\langle\psi|=0$ are $\hat{H}|\psi\rangle =  0$ and $\hat{L}|\psi\rangle = 0$ \cite{kraus_Preparation_2008}.
Here $\hat{L}$ is the entanglement-stabilizing jump term; in our scheme $\hat{L}=\hat{c}$, (cf. Eq.~\eqref{eq:closs}).
The uniqueness of this state also requires that the transition rate from the Hilbert space orthogonal to $|\psi\rangle$ to the steady state must be nonzero \cite{schirmer_Stabilizing_2010}. 
To see what in particular is required for the transition rate to be nonzero, we consider the time evolution of the overlap of a generic initial state $\hat{\rho}$ with the steady state:
\begin{align}
    \frac{{d}}{dt}\langle \psi| \hat{\rho} |\psi\rangle &= -i\langle \psi| [\hat{H},\hat{\rho}] |\psi\rangle + \langle \psi| \mathcal{D}[\hat{L}](\hat{\rho})|\psi\rangle\\
    & = \langle \psi| \hat{L}\hat{\rho}\hat{L}^\dagger|\psi\rangle.
\end{align}
If $\hat{L}^\dagger |\psi \rangle = 0$, then the transition rate from any state $\hat{\rho}$ to $|\psi\rangle$ is zero, $\partial_t \langle \psi| \hat{\rho} |\psi\rangle = 0$. In other words, the steady state is completely disconnected from its orthogonal space. 
Thus we require $\hat{L}^\dagger |\psi \rangle \neq 0$ to guarantee a unique two-qubit steady state (and thus a finite dissipative gap of the dynamics).

The origin of the vanishing dissipative gap can be traced to the conservation of total angular momentum \cite{brown_Trade_2022,doucet_High_2020}. 
Suppose we have some dissipative dynamics $\hat{\mathcal{L}}$ that stabilizes an ideal Bell state which, without loss of generality, we take to be $|S\rangle = (|01\rangle - |10\rangle )/\sqrt{2}$.
Suppose the jump operator $\hat{L}$ is restricted to linear combinations of spin operators $\hat{L} = c_1^+ \hat\sigma_1^+ + c_1^- \hat\sigma_1^- + c_2^+ \hat\sigma_2^+ + c_2^- \hat\sigma_2^-$, as is the case in the schemes of Refs.~\cite{schirmer_Stabilizing_2010,stannigel_Drivendissipative_2012,motzoi_Backactiondriven_2016,govia_Stabilizing_2022}.
The steady state condition $\hat{L} |S\rangle =0$ further confines the form of jump operator $\hat{L} = c^- \hat S^- + c^+ \hat S^+$, where $\hat{S}^- = \hat{\sigma}_1^- + \hat{\sigma}_2^-$ is the collective spin lowering operator. 
Therefore, it commutes with the total angular momentum $[\hat{L},\hat{S}^2] = [\hat{L}^\dagger,\hat{S}^2] =0$. 
Thus, total angular momentum is conserved in the jump process, and the singlet subspace is decoupled from triplet subspace, i.e.,
\begin{equation}
    \hat{L}^\dagger |S\rangle = 0.
\end{equation}
Stabilizing a perfect Bell state in finite time is thus not possible when the dissipation is a linear sum of raising and lowering operators. 
One may readily extend this argument to the remaining three Bell states via the appropriate unitary transformations.

%%%%%%%%%%%%%%%%%%%%%%%%%%%%%%%%%%%%%%%%%%%%%%%%%%%%%%%%%%%%%%%%%%%%%%
%%%%%%%%%%%%%%%%%%%%%%%%%%%%%%%%%%%%%%%%%%%%%%%%%%%%%%%%%%%%%%%%%%%%%%

\section{Strong driving steady state degeneracy}
\label{app:steady-state-degeneracy}
Here, we derive the degenerate steady state that emerges for $|\Omega/\Gamma|\to\infty$ in the $N=1$ (two-qubit) system (cf.~Eq.~\eqref{eq:qme}). For concreteness, we consider $\Delta =0$ and $\nu =1$, but the result generalizes in a straightforward manner. 
When taking strong driving limit $\Omega/\gamma\to\infty$, one cannot simply ignore the dissipation because the Hamiltonian dynamics alone never has a unique steady state (for any $\gamma>0$, the dissipation is required to pick out a unique pure steady state).

We start by (nearly) diagonalizing the Hamiltonian $\hat{H} = \hat{H}_{\rm drive} + \hat{H}_{\rm diss}$ (cf.~Eqs.~\eqref{eq:Hdrive} and \eqref{eq:Hdiss}), which can be done using a pair of equal but opposite $\pi/2$ qubit rotations about the $y$-axes,
\begin{align}
    \hat{U} = \exp\Big[\frac{i\pi}{4}\hat{\sigma}^y_{A,1}\Big]\exp\Big[-\frac{i\pi}{4}\hat{\sigma}^y_{B,1}\Big].
\end{align}
The Hamiltonian is thus
\begin{align}
    \hat H^\prime  = \hat{U}\hat{H}\hat{U}^\dagger = \frac{\Omega}{2}\left(\hat\sigma^z_{A,1} - \hat\sigma^z_{B,1}\right) + \mathcal{O}(\gamma/\Omega),
\end{align}
where we neglect the small $\gamma/\Omega\ll1$ corrections; it is safe to ignore these here because, as one can show, they will only contribute $\mathcal{O}(\gamma/\Omega)$ corrections to the final degenerate steady state.
The dissipation (cf.~Eq.~\eqref{eq:closs}) transforms as
\begin{align}
    \hat{c}^\prime = \hat U\hat c\hat U^\dagger = \frac{1}{2}(\hat \sigma^z_{A,1} - \hat\sigma^z_{B,1}) -\frac{i}{2}(\hat\sigma^y_{A,1} + \hat\sigma^y_{B,1}).
\end{align}
Now, we move into the rotating frame of the two qubits, set by the detunings $\pm \Omega/2$, and find that the dissipation has three sets of terms, each rotating at a different frequency: $0,\pm\Omega$.
We may therefore make a rotating wave approximation and split the dissipator:
\begin{align}
    \mathcal{D}[\hat{c}^\prime] \approx \mathcal{D}[\hat{L}_z] + \mathcal{D}[\hat{L}_+] + \mathcal{D}[\hat{L}_-],
\end{align}
where $\hat{L}_z = \hat{\sigma}_{A,1}^z - \hat{\sigma} _{B,1}^z$, $\hat{L}_+ = \hat{\sigma}_{A,1}^+ - \hat{\sigma}_{B,1}^-$, and $\hat{L}_- = \hat{\sigma}_{A,1}^- - \hat{\sigma}_{B,1}^+$.

Now, we find that the maximally mixed state $\mathbb{I}/4$ becomes a dark state of the dissipation up to $\mathcal{O}(\gamma/\Omega)$ corrections, and thus becomes a steady state of the dynamics.
Going back to the lab frame, we thus have the degenerate steady state
\begin{equation}
    \hat{\rho}_\nu = \frac{1}{4} (1-\nu) \hat{\mathds{I}} + \nu |S\rangle \langle S |,
\end{equation}
where $-1/3\leq \nu \leq 1$ to ensure $\hat{\rho}_\nu$ is a valid density matrix.

%%%%%%%%%%%%%%%%%%%%%%%%%%%%%%%%%%%%%%%%%%%%%%%%%%%%%%%%%%%%%%%%%%%%%%
%%%%%%%%%%%%%%%%%%%%%%%%%%%%%%%%%%%%%%%%%%%%%%%%%%%%%%%%%%%%%%%%%%%%%%

\section{Dissipative gap optimization over coupling to third level}
\label{app:optimization-eta}

%%%%%%%%%%%%%%
 \begin{figure}[t]
    \centering
    \includegraphics[width=0.78\columnwidth]{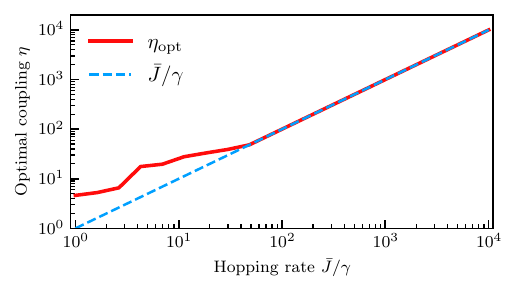}
     \caption{
        \textbf{Optimized 
        value of coupling asymmetry in the qutrit scheme.} 
        As discussed in 
        Sec.~\ref{subsce:qutrit}, replacing qubit $B1$ with a qutrit in our double chain setup allows for a dramatic acceleration of entanglement stabilization.  Here, we plot the optimized value of the coupling parameter $\eta$ (c.f.~Eq.~\eqref{eq:qutrit-master}) that minimizes the relaxation time, for each value of hopping rate $\Jbar/\gamma$.  Parameters correspond to Fig.~\ref{fig:qutrit-scheme} of the main text. For large $\bar{J}$, we find that $\eta \propto \Jbar$.
    }
     \label{fig:eta}
 \end{figure} 
%%%%%%%%%%%%%%

We provide further details here on the approach introduced in Sec.~\ref{subsce:qutrit} that lets us dramatically speed up entanglement preparation by introducing a single qutrit.  
As discussed, the entangled steady state is independent of the parameter $\eta$ in our extended master equation  Eq.~\eqref{eq:qutrit-master}, where $\eta$ is the asymmetry between the coupling of the $2-1$ and $1-0$ transitions of the $B1$ qutrit to the unidirectional waveguide.  Thus, for all other model parameters fixed, we optimize (via the Nelder-Mead method) the choice of $\eta$ to yield the smallest relaxation time (i.e. largest dissipative gap of the full Lindbladian).  The optimal choice of $\eta$ corresponding to parameters in Fig.~\ref{fig:qutrit-scheme} are shown in Fig.~\ref{fig:eta}.

 \begin{figure}[ht]
    \centering
    \includegraphics[width=0.98\columnwidth]{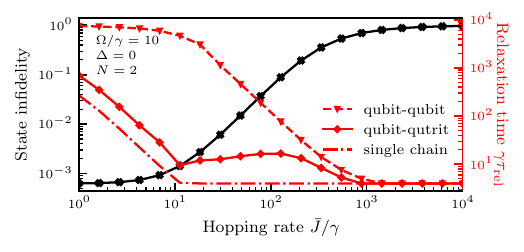}
     \caption{
        \textbf{Speeding up entanglement stabilization time of the $N=2$ nonreciprocal double chain using a qutrit.} 
        The numerically computed relaxation time $\gamma\tau_{\rm rel}$ (red) and the infidelity of the steady state to the maximally entangled state $|\psi_\infty\rangle$ (red) are shown as functions of the hopping rate $\Jbar/\gamma$ for $N=2$. Parameters are the same as those used in 
        Fig.~\ref{fig:qutrit-scheme} ($\Omega/\gamma = 10$, and $\Delta = 0$), except now $N=2$.
        For a fixed state fidelity of 0.999 (achieved at $\Jbar = 7\gamma$, dashed line), the relaxation times are $\gamma\tau_{\rm 2qb} = 4.7\times 10^3$, $\gamma\tau_{\rm qutrit} = 9.6$ and $\gamma\tau_{\rm sc} = 4.1$, for the qubit-qubit, qubit-qutrit, and single chain, respectively.
    }
     \label{fig:timescale-N2}
 \end{figure} 
 
We also present results analogous to those shown in Fig.~\ref{fig:qutrit-scheme} of the  main text, but now for a smaller system where each chain has $N=2$ sites, with $B1$ again being a qutrit.  The key features seen in Fig.~\ref{fig:qutrit-scheme}(b) of the main text are also apparent here, in particular the optimized qutrit scheme allows an order-of-magnitude speed up versus the qubit-only scheme.  

Finally, we find that the even-odd physics discussed in the main text (that emerges for weak $\drive$) also has an impact for timescales.  In Fig.~\ref{fig:timescale-N2} 
we see that for large $\Jbar$ (corresponding to small $\drive$), the relaxation time for the full double-chain system approaches the timescale for the single chain system.  In contrast, for the $N=3$ system in Fig.~\ref{fig:qutrit-scheme}, this is not the case.

\end{document}